\documentclass[journal,onecolumn]{IEEEtran}

\usepackage{amsthm}
\usepackage{latexsym}
\usepackage{exscale}
\usepackage{fontenc}
\usepackage{graphicx}

\usepackage{cite}
\usepackage{epsfig}

\usepackage{amsmath,amssymb}

\newtheorem{proposition}{Proposition}
\newtheorem{theorem}{Theorem}

\newtheorem{cor}{Corollary}

\newtheorem*{remark}{Remark}

\newcommand{\Id}{\mathbb{I}}

\begin{document}

\title{Second order statistics characterization of Hawkes processes \\
and non-parametric estimation}

\author{E.~Bacry\thanks{E.~Bacry is in the Centre de Math\'ematiques Appliqu\'ees, Ecole Polytechnique, 91128 Palaiseau, France.}
and J.F.~Muzy\thanks{J.F.~Muzy is in the
Laboratoire Sciences Pour l'Environnement, CNRS, Universit\'e de Corse,
UMR 6134, 20250 Cort\'e, France.}}

\maketitle

\begin{abstract}
We show that the jumps correlation matrix of a multivariate
Hawkes process is related
to the Hawkes kernel matrix through a system of Wiener-Hopf integral
equations.
A Wiener-Hopf argument allows one to prove that
this system (in which the kernel matrix is the unknown) possesses a unique causal solution and consequently that the second-order
properties fully characterize a Hawkes process.
The numerical inversion of this system of integral equations
allows us to propose a fast and efficient method,
which main principles were initially sketched
in \cite{BM2013}, to
perform a non-parametric estimation of the Hawkes kernel matrix. In this paper, we perform a systematic study of this non-parametric estimation procedure in the general framework of marked Hawkes processes.
We describe precisely this procedure step by step. We discuss the estimation error and explain how the values for the main parameters should be chosen.
Various numerical examples are given in order to illustrate the broad possibilities of this estimation procedure ranging from 1-dimensional (power-law or non positive kernels) up to 3-dimensional
(circular dependence) processes. A comparison to other non-parametric estimation procedures is made.
Applications to high frequency trading events in financial markets and to earthquakes occurrence dynamics are finally considered.
\end{abstract}


\section{Introduction}
A (multivariate) Hawkes process is a counting process which intensity, at each time,
is given by a linear regression over past jumps of the process \cite{Hawkes1971-1,Hawkes1971-2}.
This ``self-'' and ``mutually-'' exciting nature
of Hawkes processes makes them very appealing to account, within a simple
tractable model, for situations where the
likelihood of future events directly depends on the occurrence of past events.
For this reason, there has been, during last decade, a growing interest for Hawkes processes
in various fields where endogenous triggering, contagion, cross-excitation can be naturally
invoked to explain discrete events dynamics. Originally Hawkes models have been introduced to
describe the occurrence of earthquakes in some given region \cite{ogata99,helsor02}, but they also became popular in many other areas like high-frequency finance (trading and order book events dynamics) \cite{BH,Hewlett2006Clustering,BDHM2010}, neurobiology (neurons activity) \cite{Rey2}, sociology (spread of terrorist activity) \cite{emhawkes,mo11} or processes on the internet (viral diffusion across social networks) \cite{cranesor08,yangzha}.

A $D$-dimensional Hawkes process $N(t)$ is mainly
characterized by its $D \times D$ kernel matrix $\Phi(t)=\{\phi^{ij}(t)\}_{1\le i,j\le D}$, where the kernel $\phi^{ij}(t)$  describes how events of the $j$th process  component
influence the occurrence intensity of the $i$th  component.
As far as estimation of Hawkes process is concerned, in most studies, one assumes
a specific parametric shape for the kernel components $\phi^{ij}(t)$ (e.g. an exponential decay)
and performs either a moment method (based e.g. on the bartlett spectrum \cite{bartlett_spectral_1963,bartlett_spectral_1964})
or a maximum likelihood estimation \cite{ozaki_maximum_1979,ogata_linear_1982}.
However, when one has no a priori on the shape of the kernel components, it is necessary
to perform a non-parametric estimation. The first approach devoted to this goal was
a method based on Expected Maximization (EM) procedure of a (penalized) likelihood function \cite{MaLe10,emhawkes}.
It has been designed for monovariate Hawkes processes and can be hardly used to handle large amounts of data in situations where the kernel function is not well localized as compared to the exogenous inter-events time (see section \ref{otherapproaches}).
Another approach, proposed in \cite{Rey0,Rey1,Rey2}, consists in minimizing a contrast function,
assuming that the kernels  $\{\phi^{ij}(t)\}_{1\le i,j\le D}$ can be decomposed over atoms of some dictionary. The estimation is regularized using  a Lasso penalization which provides sparse estimations.
The so-obtained kernels are, for instance,  piece-wise constant functions with very few non zero pieces.
Though it is clearly the right choice when the dimension $D$ is large or when the kernels are known to be very well localized (this seems to be the case when modeling networks of neurons \cite{Rey2}), it does not make sense when modeling earthquakes or financial time-series for which a large amount of data is available and the kernels are known to be power-law. Let us also mention an approach proposed in \cite{BaMuDaEPJB}, where the authors estimate the kernel
components $\phi^{ij}(t)$ from the jumps correlation matrix through a spectral method. This technique
is however exclusively adapted to symmetric Hawkes processes.

In this paper, our main purpose is two-fold: We first provide a complete
overview of Hawkes processes second order properties and show that
they uniquely characterize the process, i.e., there is a one-to-one
correspondence between the matrix $\Phi(t)$ and a matrix $g(t)$ associated with the jumps ($dN(t)$)
correlation function. In fact, we prove that $\Phi(t)$ is the unique causal solution of
a system of Wiener-Hopf equations that directly involves the correlation function.
We then show that
this property allows us to propose a new simple non-parametric
estimation method based on the explicit resolution
of the Wiener-Hopf system using a Gaussian quadrature method.
In many applications, it is interesting to consider a marked version of Hawkes processes,
where the conditional intensities of the Hawkes components may also depend on some random
variables (the marks) associated with each event through a matrix of mark functions.
We write the new system of equations which links $\Phi(t)$ with the correlation function which is no longer a Wiener-Hopf system. In the case the mark functions are piecewise constant, we show that a simple variant
of our method allows one to solve this system, i.e., to recover both the kernel and the mark functions matrices.
We provide various numerical examples ranging from a 1-dimensional process up to a 3-dimensional process.
We also provide
some heuristics about the estimation error
and convergence issues as well as some procedures to fix the various parameters of the method.
We compare this method to existing non-parametric methods and show that it is relatively simple to implement and allows one to handle very
large data sets and lead to reliable
results in a wide number of situations.
Our estimation framework is then illustrated on two different applications:
First, along the same line as in Refs. \cite{BaMuDaEPJB,BouchaudEPJB2013},
we study the arrival of market orders of 2 liquid future assets on the EUREX exchange.
We show that the estimated kernel is strikingly well fitted by a power-law function with an
exponent very close to $1$. Second, we estimate a Hawkes model of earthquake events marked
by the event magnitudes from the Northern California EarthQuake Catalog (NCEC) \cite{ncedc}.
To our knowledge, the only attempt to estimate a non-parametric self-exciting from earthquakes
was provided in Ref. \cite{MaLe10}. In our case we confirm that the ETAS model specifications, i.e.
of power-law shape of the kernel $\phi(t)$ and an exponential law for the mark function.

The paper is structured as follows: In section \ref{Sec:MultivariateHawkesProcesses}, we introduce the main mathematical definitions and review the second order properties of multivariate Hawkes processes.
We notably provide the explicit expression for the Laplace transform of the jump correlation function.
We also introduce the conditional expectation matrix $g(t)$ and
show how it is basically related to the jump correlation function.
Finally, we recover a former result by Hawkes \cite{Hawkes1971-2}
by virtue of which the kernel matrix $\Phi(t)$
satisfies a Wiener-Hopf equation with $g(t)$ as a Wiener-Hopf kernel.
We show that the   Wiener-Hopf operator has a unique causal solution and thus
$\Phi(t)$ is uniquely determined
by the shape of $g(t)$.
Our estimation method relies on this property and is described in section \ref{estimationmethod}. More precisely,
by reformulating a marked Hawkes process with piecewise constant mark functions as a multivariate Hawkes process
in higher dimension, we propose a numerical method that allows one to estimate both the kernel and the mark functions.
This method mainly uses a Nystr\"om method to solve the system of Wiener-Hopf equations.
It is illustrated in Section \ref{sec:num} on four different examples : a 2-dimensional marked process, two 1-dimensional processes (one involving slowly decreasing kernels and one involving a negative-valued kernel) and a 3-dimensional process (involving circular dependance and non decreasing kernels).
Section \ref{error_analysis}  presents some heuristics about the analysis of the estimation errors and
discusses some procedures to fix the main parameters of the method (namely, the bandwidth and the number of quadrature points).
We then briefly discuss the link to  former approaches for non-parametric kernel estimation  in section
\ref{otherapproaches}, while the aforementioned applications to financial statistics and
geophysics are considered in section \ref{sec:applications}.
Conclusion and prospect are given is section \ref{sec:conclusion}.

\section{Multivariate Hawkes Processes and their second order properties \label{Sec:MultivariateHawkesProcesses}}

\subsection{The framework}
\label{framework}
We consider a $D$-dimensional point process $N_t = \{N^i_t\}_{1\le i \le D}$. Each component $N^i_t$ is a 1d point process whose jumps are all of size 1, and whose intensity at time $t$ is $\lambda^i_t$. Thus the intensity vector of $N_t$ is $\lambda_t = \{\lambda^i_t\}_{1\le i \le D}$.

We consider that $N_t$ is a Hawkes process \cite{Hawkes1971-1}, i.e., that, at time $t$, each intensity component $\lambda^i_t$ can be written  as a linear combination of past jumps of $N_t$, i.e.,
\begin{equation}
\label{lambdai}
\forall i \in [1,D],~~\lambda^i_t = \mu^i + \sum_{j=1}^D \int_{(-\infty,t)} \phi^{ij}(t-s) dN^j_s,
\end{equation}
where
\begin{itemize}
\item  $\mu =  \{\mu^i\}_{1\le i \le D}$ are exogenous intensities
\item  each  kernel function $\phi^{ij}(t)$ is a positive  and causal (i.e., its support is included in $\mathbb{R}^+$). It codes the influence of the past jumps of $N^j$ on the current intensity $\lambda^i_t$. In the following, the kernel matrix will denote the $D\times D$ matrix function
$\Phi(t)=\{\phi^{ij}(t)\}_{1\le i \le D}$.
\end{itemize}
Using matrix notations, the $D$ equations \eqref{lambdai} can be rewritten in a very synthetic way as
\begin{equation}
\label{lambda}
\lambda_t = \mu +  \Phi \star dN_t,
\end{equation}
where $\lambda =  \{\lambda^i\}_{1\le i \le D}$ and the operator $\star$ stands for regular matrix multiplications where all the multiplications are replaced by convolutions.

Let us remind that the process $N_t$ has asymptotically stationary increments (and the process $\lambda_t$ is asymptotically stationary) if
the following hypothesis holds  \cite{Hawkes1971-1} :
\begin{itemize}
\item[(H)] the matrix $||\Phi|| = \{||\phi^{ij}||\}_{1\le i \le D}$  has a spectral radius $\rho$ strictly smaller than 1,
\end{itemize}
where $||g||$ stands for $||g|| = \int g(t) dt$.
In the following we will always consider that (H) holds and we shall consider that $N_t$ corresponds to the asymptotic limit, i.e., $N_t$ has stationary increments and $\lambda_t$ is a stationary process.

In that case,  the vector $\Lambda$ of mean event rates
associated with each component reads \cite{Hawkes1971-1}:
\begin{equation}
\label{Lambda}
\Lambda = E(\lambda_t) = (\Id-||\Phi||)^{-1} \mu.
\end{equation}
In the following, $\Lambda^i$ will refer to the $i$th component of $\Lambda$.

\begin{remark}[Inhibitory effects in Hawkes processes]
\label{nlHawkes}
Let us point out that Br\'emaud and Massouli\'e \cite{bremaud1996stability} have shown
that the stability condition criterium remains valid when $\lambda_t$ is a nonlinear positive Lipschitz function of $\Phi \star dN_t$ and where the kernels $\phi^{
ij}$ are not restricted to be positive. A particular interesting generalization of Eq. \eqref{lambda} (notably considered in \cite{Rey1}) is:
\begin{equation}
\label{lambda2}
\lambda_t= \left(\mu +  \Phi \star dN_t \right)_+
\end{equation}
where $(x)_+ = x$ if $x>0$ and $(x)_+ = 0$ otherwise.
This extension allows one to account for inhibitory effects when $\phi^{ij}(t)$ can take negative values.
\end{remark}

\subsection{Characterization of a multivariate Hawkes process through its second-order statistics}
In this section we recall the main properties of the correlation matrix function associated with
multivariate Hawkes processes and prove that it fully characterizes these processes.

The second-order statistics are summed up by the infinitesimal covariances :
\begin{equation}
\label{cov}
Cov(dN^i_t,dN^j_{t'}) = E(dN^i_tdN^j_{t'}) - \Lambda^i \Lambda^j dt dt',~~~1\le i,j \le D.
\end{equation}
As already explained, under assumption (H), $N_t$ can be considered to have stationary increments, consequently,
$Cov(dN^i_t,dN^j_{t'})$ only depends on $t'-t$.
Moreover, these measures are non singular measures, except for $i=j$ for which it has a Dirac component $\delta(t)$.
Thus the non-singular part of this covariance can be written as
\begin{equation}
\label{nsp}
\nu^{ij}(t'-t) dt dt'  = E(dN^i_t dN^j _{t'}) - \Lambda^i \Lambda^j dtdt' - \Lambda^{i} \epsilon_{ij}\delta(t'-t)dt,~~~1\le i,j \le D,
\end{equation}
where $\epsilon_{ij}$ is the Kronecker symbol which is always 0 except when $i=j$ in which case it is equal to 1.

Let us point out that, since all the jumps of the process $N$ are of size 1, the second order statistics can be rewritten in terms of conditional expectations (see \cite{BM2013}). Indeed, for all  $1\le i,j \le D$, let $g^{ij}(t)$ be the non-singular part of the density of the measure  $E(dN^i_t - \Lambda^i dt| dN^j_0 = 1)$, i.e.,
\begin{equation}
\label{gij}
g^{ij}(t) dt = E(dN^i_t| dN^j_0 = 1) - \epsilon_{ij} \delta(t) -\Lambda^i dt.
\end{equation}
Then
\begin{eqnarray*}
\nu^{ij} (t'-t) dt dt'& = & E(dN^i_t dN^j_{t'}) - \Lambda^i  \Lambda^j  dtdt' - \Lambda^i \epsilon_{ij}\delta(t'-t)dt\\
& = & E(dN^j_{t'} | dN^i_t=1)Prob\{dN^i_t = 1) - \Lambda^i  \Lambda^j  dtdt'  - \Lambda^i \epsilon_{ij}\delta(t'-t)dt\\
& = & E(dN^j_{t'-t} | dN^i_0=1)\Lambda^i dt - \Lambda^i  \Lambda^j  dt dt'  - \Lambda^i \epsilon_{ij}\delta(t'-t)dt\\
& = & \Lambda^i g^{ji}(t'-t)dt dt'
\end{eqnarray*}
It follows that:
\begin{equation}
\label{relnug}
\nu (t) = \Sigma g^T (t)
\end{equation}
where $g^T(t)$ stands for the transpose of the matrix $g(t)$.

In \cite{BaMuDaEPJB}, it has been shown that
the "infinitesimal" covariance matrix $\nu(t)=\{\nu^{ij}(t)\}_{0\le i,j < D}$ can be directly
related to the kernel matrix $\Phi$ :
\vskip .3cm
\noindent
\begin{proposition}[from \cite{BaMuDaEPJB}]\label{prop:InfinitesimalCov}
Let $\Psi(t) = \sum_{k=1}^{+\infty} \Phi^{(\star k)}(t)$, where $\Phi^{(\star k)}$ stands for the matrix convolution
$\Phi \star \Phi \star \ldots \star \Phi$ (where $\Phi$ is repeated $k$ times).
Let  $\widetilde{\Psi}(t)=\Psi(-t)$.
Then
\begin{equation}
    \nu(t) =
       (\delta\Id + \widetilde \Psi) \star \Sigma (\delta\Id +\Psi^T)(t)  -\delta(t) \Sigma
\label{eq:EdNdNPrime}
\end{equation}
where $\Sigma$ is the diagonal matrix defined by $\Sigma^{ii} = \Lambda^i$ and where
we use the convention
$\delta \star \delta(t) = \delta(t)$.
In the Laplace domain, this last equation writes
\begin{equation}
    \hat \nu(z) =
       (\Id + \hat \Psi(-z)) \Sigma (\Id + \Psi^T(z))  - \Sigma.
\label{eq:lEdNdNPrime}
\end{equation}
Moreover,
\begin{equation}
\label{lpsi}
\Id + \hat \Psi(z) = (\Id - \hat \Phi(z))^{-1},
\end{equation}
where the Laplace transform of a $L^1$ causal function $s(t)$ is defined as:
\begin{equation}
\hat s(z) = \int_{-\infty}^{+\infty} s(t) e^{zt} dt,~~\forall z \in \mathbb{C} \; .
\end{equation}
\end{proposition}
\vskip .3cm
\noindent
From Eqs \eqref{relnug} and \eqref{eq:EdNdNPrime}, it results:
\begin{equation}
\label{g}
g(t)=  \nu^T (t) \Sigma^{-1} =  (\delta\Id + \Psi) \star \Sigma (\delta\Id + \widetilde \Psi^T)(t)\Sigma^{-1}   - \delta(t)\Id.
\end{equation}
or alternatively, in the Laplace domain
\begin{equation}
\label{lg}
\hat g(z)=    (\Id + \hat \Psi(z)) \Sigma (\Id + \hat \Psi^T(-z))\Sigma^{-1}   - \Id.
\end{equation}
Since
 \begin{equation}
 \label{psi}
 (\delta\Id-\Phi) \star (\delta\Id+\Psi)(t) =  \delta(t) \Id,
 \end{equation}
convoluting both hand sides of \eqref{g} by $\delta(t)\Id-\Phi(t)$ leads to
\begin{equation}
(\delta \Id-\Phi) \star g(t)   =
 \Sigma (\delta(t) \Id + \widetilde \Psi^T(t))\Sigma^{-1}   - \delta(t)  \Id+\Phi(t)
\end{equation}
which gives
\begin{equation}
(\delta\Id-\Phi) \star g(t)  = \Sigma\widetilde \Psi^T(t) \Sigma^{-1}  +\Phi(t)
\end{equation}
Since all the elements of $\Phi(t)$ are causal functions (i.e., supported by $\mathbb{R}^+$) and since $\Psi(t)$ is just expressed as convolution products of the matrix $\Phi$, all the elements of $\Psi(t)$ are also causal functions.
Writing this last equation  for $t>0$, thus leads to the following Proposition:



\vskip .3cm
\noindent
\begin{proposition}\label{prop:WienerHopf}
The kernel matrix functions $\Phi(t)$ statisfies the $D^2$-dimensional Wiener-Hopf system
\begin{equation}
\label{wh}
g(t) = \Phi(t) + \Phi \star g(t),~~~\forall t >0
\end{equation}
\end{proposition}
Wiener-Hopf equations have been extensively studied in the past century \cite{WH,GK1958,Noble} and
arise in wide variety of problems in physics.
This last system of equation was first established by Hawkes in \cite{Hawkes1971-2} (see also \cite{BM2013}) in order to express $g(t)$ (considered as the "unknown") as a function of $\Phi(t)$.
He proved that it has a unique solution in $g(t)$.
One can work the other way around and consider it as a system of equations that $\Phi(t)$ (the "unknown") satisfies, $g(t)$ being known.
In Appendix \ref{sec:unicity}, thanks to the famous factorization
technique introduced by Wiener and Hopf \cite{Noble},
we prove that this system  \eqref{wh} admits a unique solution in $\Phi(t)$:
\vskip .3cm
\begin{theorem}
\label{theorem}
Let $N_t$ be a $D$-dimensional Hawkes process with exogenous intensity $\mu$ and kernel matrix $\Phi(t)$ as defined in Section \ref{framework}. Let $\Phi(t)$ satisfies hypothesis (H) (as explained above, we then can consider that $N_t$ has stationary increments).
Let $g(t)$ be the matrix $g(t)=\{g^{ij}(t)\}_{1\le i,j\le D}$ defined by \eqref{gij}.
Then the matrix $\Phi(t)$ is the only solution (whose elements are all causal and in $L^1(\mathbb{R}^+)$) of the system of equations (in which $\chi(t)$ is the unknown)
\begin{equation}
g(t) = \chi(t)+\chi(t) \star g(t),~~\forall t >0.
\end{equation}
\end{theorem}
One can then state  the corollary :
\vskip .3cm
\begin{cor}
A multi-dimensional Hawkes process with stationary increments is uniquely defined by
its first-order statistics (i.e., the expectation of its intensity $\lambda(t)$) and its second-order statistics
(given by either its correlation function \eqref{cov} or, equivalently, by its conditional expectations \eqref{g}).
\end{cor}

\vskip .2cm
\noindent
Indeed, this is a direct consequence
\begin{itemize}
\item of  the Theorem \ref{theorem} which proves that the second-order statistics fully characterize the kernel matrix $\Phi(t)$
\item and of  \eqref{Lambda} which allows to express $\mu$ as a function of $||\Phi||$ and the first-order statistics $\Lambda$.
\end{itemize}

\section{Non-parametric estimation of a multi-dimensional marked Hawkes process : framework and main principles}
\label{estimationmethod}
As we shall see at the end of this Section, the results of previous Section naturally lead to a non-parametric estimation method for multidimensional Hawkes processes.
Actually, we shall develop our estimation framework in a slightly more general framework
than the one presented above which is very useful in applications : the framework of multi-dimensional marked Hawkes process.


\subsection{The marked Hawkes process framework}
\label{mframework}
Using the same notation as in Section \ref{framework},
each component $N^i_t$ is now associated with observable iid marks $\xi^i_t$.
$\xi^i_t$ is non zero only at time when $N^i_t$ jumps. In the following, we note
$p^i(x)$ the density of the law of the random variable $\xi^i_t$ conditionally to the fact $N^i_t$ jumps.
We then replace \eqref{lambdai} by the new equation
\begin{equation}
\label{mlambdai}
\forall i \in [1,D],~~\lambda^i_t = \mu^i + \sum_{j=1}^D \int_{(-\infty,t)} \phi^{ij}(t-s) f^{ij}(\xi^j_s)dN^j_s,
\end{equation}
where
the past  observed marks $\xi^j_s$
 influence the current intensity $\lambda^i_t$
through the unobserved mark (positive) function $f^{ij}(x)$. Since $f^{ij}(x)$ is only involved through the product $\phi^{ij}(t)f^{ij}(\xi^j)$, it is defined up to a multiplication factor. We fix this factor by choosing $f^{ij}$ such that $E(f^{ij}(\xi^j)) = 1$.

As in the unmarked case, it is easy to show that under the condition (H), one can consider that $N_t$ has stationary increments and that \eqref{Lambda} still holds.

In place of the functions $g^{ij}(t)$ for an non-marked Hawkes process (see \eqref{gij}),
we now define the function $G^{ij}(t,x)$ as
\begin{equation}
\label{defGij}
G^{ij}(t,x)dtdx = E(dN^i_t~ |~ dN^j_0=1 ~\mbox{and}~\xi^j_0 \in [x,x+dx]) - \epsilon_{ij}\delta(t)dx -\Lambda^i dtdx.
\end{equation}
Let us note that $g^{ij}(t) = \int G^{ij}(t,x) p^i(x) dx$.
In Appendix \ref{app:proofinteqmarked} we prove that $G^{ij}$ satisfies a system
of integral equation as stated by the following proposition:
\vskip .3cm
\begin{proposition}
\label{prop:inteqmarked}
Let $i \in [1,D]$ fixed. The kernels $\{\phi^{ij}(t)\}_{1\le j \le D}$ of the marked Hawkes process defined in Section \ref{mframework} satisfy the following system of integral equations~:
\begin{equation}
\label{msystem}
\forall j \in [1,D],~ \forall t > 0, ~\forall x,~~G^{ij}(t,x) =  \phi^{ij}(t)f^{ij}(t) +
\sum_{k=1}^D   \phi^{ik} \star K_x^{ijk}(t),
\end{equation}
where
\begin{equation}
\label{mk}
K_x^{ijk}(t) =
G^{kj}(t,x) 1_{\mathbb{R}^{+}}(t)
+
\frac{\Lambda^k }{\Lambda^j}
\int dz  f^{ik}(z)
p^k(z) G^{jk}(-t,z)1_{\mathbb{R}^{-}}(t).
\end{equation}
\end{proposition}

\subsection{The Wiener-Hopf equation in the case of piece-wise constant $f^{ij}$ mark functions}
In a non-parametric estimation framework, the unknown of the system \eqref{msystem} are both the kernels $\phi^{ij}$ {\em and} the mark functions $f^{ij}$.
In the case of non-marked Hawkes process, we already showed (see Eq. \eqref{wh}) that this system is a Wiener-Hopf system.
We proved the unicity of the solution and, one can use standard methods for solving it.
The system \eqref{msystem} is no longer a Wiener-Hopf system which makes things much harder.

In the particular case of piece-wise constant mark functions $f^{ij}$ (with $M$ number of pieces), it is easy to prove that the process $N$ basically
corresponds to a non-marked Hawkes process of dimension $D M$ : each components corresponds to the jumps of a component of
the marked Hawkes process associated with one of the $M$ mark function values. Thus it is clear that the system \eqref{msystem} can be written in terms of a $D^2M$ Wiener-Hopf system of the type of \eqref{wh}. It will be the base of our non-parametric estimation procedure.

\vskip .3cm
\noindent
{\bf Remark : } Before moving on, let us point out that,  in the  case the marks $\xi^{ij}$
influence the intensity of the process only through a finite number $M$ of values, one could consider the more general framework where the kernels themselves depend on the marks, i.e.,
\begin{equation}
\forall i \in [1,D],~~\lambda^i_t = \mu^i + \sum_{j=1}^D \int_{(-\infty,t)} \tilde \phi^{ij}(t-s,\xi^j_s)dN^j_s,
\end{equation}
and, following the same argument as above, estimate all the kernels $\tilde \phi^{ij}(.,.)$
solving a $D^2M$ Wiener-Hopf system.
This can be achieved following the exact same lines as the algorithm described in the next Section. For the sake of simplicity,
we will describe the algorithm in the case $\tilde \phi^{ij}(t,\xi) = \phi^{ij}(t)f^{ij}(\xi)$.

We consider that for any $j$, there exists a covering of $\mathbb{R}$ with a finite number of intervals $\{I^j(l)\}_{1\le l\le M^j}$ such that, for any $i$ and for any $l$,
$f^{ij}(x)$ restricted to $x\in I^j(l)$ is a constant function which value is $f^{ij}_l$~:
\begin{equation}
\forall x \in I^j(l), ~f^{ij}(x) = f^{ij}_l.
\end{equation}
Since the  marks $\xi^{ij}$ are involved in the process construction only through the functions  $f^{ij}(\xi^{ij})$, it is clear that the functions $x \rightarrow G^{ij}(t,x)$ are also piece-wise constant.
Thus we note
\begin{equation}
\forall x \in I^j(l), ~G^{ij}(t,x) = G^{ij}_l(t).
\end{equation}
In the same way, we note
\begin{equation}
\label{plj}
\int_{I^j(l)} p^j(x) dx = p^{j}_l.
\end{equation}
For fixed $i$ and $j$, and for a fixed $x\in I^j(l)$, Equations \eqref{msystem} and \eqref{mk} then rewrite
\begin{equation}
\forall t > 0, ~~G^{ij}_l(t) =
\phi^{ij}(t) f^{ij}_l(t)
+
\sum_{k=1}^D \phi^{ik} \star \left(G^{kj}_{l}(t)1_{\mathbb{R}^+}(t)
+
\frac{\Lambda^k }{\Lambda^j}
\sum_{m=1}^{M^k}
 f^{ik}_m
p^k_m G^{jk}_{m}(-t)1_{\mathbb{R}^-}(t)\right)
\end{equation}
Since, $E(f^{ik}(\xi^{ik})) = 1$, one has $\sum_{m=1}^{M^k}f^{ik}_mp^{k}_m = 1$, then, if we make the change of variable $\phi^{ik}_m(t) = f^{ik}_m \phi^{ik}(t)$ (consequently
$\sum_{m=1}^{M^k} p^{k}_m\phi^{ik}_m(t) = \phi^{ik}(t)$), we get
\begin{equation}
\label{system1}
\forall t > 0, ~~G^{ij}_l(t) =
\phi^{ij}_l(t)
+ \sum_{k=1}^D
\sum_{m=1}^{M^k}
\phi^{ik}_m \star \left( p^{k}_mG^{kj}_{l}(t) 1_{\mathbb{R}^+}(t)
+
\frac{\Lambda^k }{\Lambda^j}
p^k_m G^{jk}_{m}(-t)1_{\mathbb{R}^-}(t)\right)
\end{equation}
We finally get the Wiener-Hopf system

\vskip .3cm
\begin{proposition}
Let $i \in [1,D]$ fixed. The marked kernels $\{\phi^{ij}_l(t)=f^{ij}_l\phi^{ij}(t)\}_{1\le j \le D,1\le l \le M_j}$  satisfy the following Wiener-Hopf system~:
\begin{equation}
\label{msystem1}
\forall t > 0, ~~G^{ij}_l(t) =
\phi^{ij}_l(t)
+ \sum_{k=1}^D
\sum_{m=1}^{M^k} \int_0^{+\infty}  \phi^{ik}_m(s) K_l^{ijk}(t-s) ds,
\end{equation}
where
\begin{equation}
\label{mk1}
K_l^{ijk}(t) = p^{k}_mG^{kj}_{l}(t) 1_{\mathbb{R}^+}(t)
+
\frac{\Lambda^k }{\Lambda^j}
p^k_m G^{jk}_{m}(-t)1_{\mathbb{R}^-}(t).
\end{equation}
Moroever,
\begin{equation}
\phi^{ik}(t) = \sum_{m=1}^{M^k} p^{k}_m\phi^{ik}_m(t)
\end{equation}
\end{proposition}
Thus we get $D$ Wiener-Hopf systems of dimension $DM$ where $M=\sum_{j=1}^DM^j$.
The unicity of the solution is actually deduced from the result of Section \ref{sec:unicity} where we proved the
unicity of the solution of the system \eqref{wh}. Indeed, as we explained in the previous Section, the marked Hawkes process we considered here is equivalent to a non-marked Hawkes process of dimension $DM$.

Let us point out that, in the case the functions $f^{ij}$ are constant functions (and thus equal to 1, since $E(f^{ij}(\xi^{ij}_t)) = 1$),
then we recover the system \eqref{wh} (which corresponds to $D$ Wiener-Hopf systems of dimension $D$).
 Indeed, in that case we have $G^{ij}_l(t) = g^{ij}(t)$,  $\phi^{ij}_l(t) = \phi^{ij}(t)$, $p_l^j = 1$ and, for all $t<0$,  $g^{ij}(t) = \frac {\Lambda^i}{\Lambda^j} g^{ij}(-t)$.

\subsection{Discretizing the system \eqref{msystem} using Nystr\"om method}
\label{nsmeth}
Solving the Wiener-Hopf  systems \eqref{msystem1}  is the main principle of our estimation procedure.
There is a huge literature about numerical algorithm for solving Wiener-Hopf equations \cite{Noble,At76}.
One algorithm which is particularly simple and which works particularly well is the Nystr\"om method\cite{N30}.
It basically consists in
replacing convolutions (in time) by discrete sums using quadrature method with $Q$ points.
Then, for each $i$, solving \eqref{msystem1} amounts in solving a linear $QDM$
dimensional standard linear system which can be solved by simply inverting the corresponding $QMD \times QMD$ matrix.
Of course, in order to get the full estimation of the kernels $\phi^{ij}$ and the constants $f^{ij}_l$, one has to solve $D$ such systems (one for each value of $i$). Let us point out that, when the kernels have been estimated on the quadrature points, using quadrature formula, one can compute these estimations on any finer grid.

If we suppose that the kernels are supported by $[0,A]$ one gets
\begin{equation}
\label{system3}
\forall t > 0, ~~G^{ij}_l(t) =
\phi^{ij}_l(t)
+ \sum_{k=1}^D
\sum_{m=1}^{M^k} \int_0^A \phi^{ik}_m(s) K_l^{ijk}(t-s) ds.
\end{equation}
However, when using the Nystr\"om method, i.e., computing the integral using quadrature methods, one has to be careful since  the kernel $K_l^{ijk}(t)$ is generally discontinuous at $t=0$.
A standard way to deal with a Wiener-Hopf kernel which is singular (non exploding) on the diagonal, is to rewrite \eqref{system3} in the following way :
\begin{equation}
\label{system4}
\forall t > 0, ~~G^{ij}_l(t) =
\phi^{ik}_l(t)
+ \sum_{k=1}^D
\sum_{m=1}^{M^k} \int_0^A (\phi^{ik}_m(s) -\phi^{ik}_m (t)) K_l^{ijk}(t-s) ds + \phi^{ik}_m (t) \int_0^A K_l^{ijk}(t-s) ds.
\end{equation}
The term in the first integral is then no-longer discontinuous (quadrature error will be one order smaller) and the second term can be estimated directly, estimating directly the primitive function of the $G^{ij}_l(t)$ functions and using the formula
\begin{equation}
\int_0^A K_l^{ijk}(t-s) ds = p_m^k \left( \int_0^{t} G_l^{kj}(s)ds +  \frac{\Lambda^k }{\Lambda^j}  \int_0^{A-t} G_m^{jk}(s) ds \right).
\end{equation}

\subsection{Main steps od our non-parametric estimation}
\label{algo1}
We are now ready to present the main steps of our estimation algorithm.
We refer the reader to Section \ref{algo} for a more detailed description of the algorithm.
The main steps for non-parametric estimation of multi-dimensional Hawkes processes are :
\begin{itemize}
\item For all $j\in[1,D]$, fix a priori the $M^j$ intervals $\{I^j(l)\}_{1\le l \le M^j}$ on which the
mark function $f^{ij}(x)$ is considered as constant.
\item For all $i,j\in[1,D]$ and $l \in [1,M]$,
estimate the conditional-law functions $G_l^{ij}(t) = G_l^{ij}(t,x \in I^j_l)$ as defined by \eqref{defGij} by replacing the expectation by empirical averages. These functions are
are density measures, in particular they are positive (cf \eqref{g}) though they do not sum to 1.
It is very natural to use classical kernel-based non-parametric estimation techniques used for estimating the density of a random variable from iid realizations \cite{P62}. This classically involves a bandwidth parameter $h$ which corresponds to the size of the support of the kernel. This well be developed in the Section \ref{Sec:bandwidth}.
\item Solve, in $\Phi$, the  Wiener-Hopf system \eqref{msystem} using the Nystr\"om method as explained in Section \ref{nsmeth} using $Q$ Gaussian quadrature points.
\item Using quadrature formula, the so-obtained estimation of $\Phi$ (at the quadrature points) can be re-sampled on any grid with an arbitrary resolution.
\end{itemize}
This method mainly involves two key parameters : the bandwidth $h$ and the number $Q$ of quadrature points.
The next section gives some insights on how one should fix these two parameters.

\section{Selection of the bandwith $h$ and the number $Q$  of quadrature points}
\label{error_analysis}
The goal of this Section is to give qualitative arguments for controlling the error of the estimation procedure presented previously
and to understand how to choose the estimation parameters $h$ (the bandwidth) and $Q$ (the number of quadrature points).
It will be illustrated by numerical simulations.

For the sake of simplicity, we shall consider that the  Hawkes process is not marked.
The arguments are exactly the same in the case of a marked Hawkes process.

There are mainly two sources of errors :
\begin{itemize}
\item[(${\cal E}_1$)] the first one comes from the estimation of the
conditional expectation density  $g^{ij}(t)$ \eqref{gij}. This error is basically controlled by the amount of data available, i.e., by the number $J$ of jumps available, and by the value of the bandwidth parameter $h$.
 \item[(${\cal E}_2$)] The other one comes from the inversion of the Wiener-Hopf system \eqref{wh}. This error is basically controlled by the number $Q$ of quadrature points.
 \end{itemize}


\noindent
In the error analysis,
it is very hard to take into account the fact that the kernel of the Wiener-Hopf system is a random function (i.e., the empirical
estimation of the conditional expectation). This problem has somewhat already been addressed in \cite{Rey1} in a particular case  (see Section \ref{otherapproaches}).
Clearly, this problem  should be solved for understanding deeply the performance of our estimator.
However, in the general case, this is a very difficult problem which is still open as of today.

%
%

%

 \vskip .5cm
\subsection{Conditional expectation density estimation - Bandwidth ($h$) selection}
\label{Sec:bandwidth}
One of the main step of the algorithm presented in Section \ref{algo1} consists in estimating the functions $g^{ij}(t)$\footnote{In the non marked case the functions
$G^{ij}_l(t)$ are such that  $G^{ij}_l(t) = g^{ij}(t)$, $\forall t$}
defined by \eqref{gij},
for $t >0$ (for negative $t$, one can use the formula $g^{ij}(-t) = g^{ji}(t) \Lambda_i \Lambda_j^{-1}$).

These functions are density measures, in particular they are positive (cf \eqref{g}) though they do not sum to 1.
It is very natural to use classical kernel-based non-parametric estimation techniques used for estimating the density of a random variable from iid realizations \cite{P62}.

Let $K(x)$ be a {\em kernel function} of order $l$, i.e., a localized function such that $\int K(u) du = 1$, $\int K^2(u) du < +\infty$
and $\int u^nK(u) du = 0$ for all  $n$, $1 \le n \le l$. Since we want to perform estimations only for $t>0$, it is convenient to choose the support
of $K(u)$ to be in $\mathbb{R}^+$.
For the sake of simplicity, we consider that we have $R$ iid realizations $\{N_t(r)\}_{1\le r \le R}$ of the Hawkes process on an interval $[0,t_{max}]$ (such that
$g_{[t_{max},+\infty[}  \simeq 0$).
For each realization $r$, we call $t^j_n(r)$ the $n$th jumping time
of the $j$th component $N^j_t(r)$.
For a given {\em bandwidth} $h>0$, it is natural to consider the following estimator of $g^{ij}(t)+\Lambda^i$ for $t > 0$
\begin{equation}
\label{averaged}
g_*^{ij}(t) = \frac{1}{Rh} \sum_{r=1}^{R} \int_0^{t_{max}}   dN^i_u(r) K\left(\frac{u-t^j_1(r)-t}h\right)
\end{equation}

The mean square error is defined as:
\[
 MSE(h)  = E({\cal E}_1^2) = E \left((g^{ij}(t) - g_*^{ij}(t))^2\right) =  Var(g_*^{ij}(t)) + b^2(t)
\]
where we denoted the estimation bias $b(t) = g^{ij}(t)-E(g_*^{ij}(t))$.
The following classical bound of this error is proven in Appendix \ref{app:esterror}:

\vskip .3cm
\noindent
\begin{proposition}[{\bf Convergence speed}]
\label{propmse}
If $g^{ij}(t)$ is H\"older $\beta$ and if the kernel $K$ is of order $l \geq \lfloor \beta -\epsilon\rfloor$, then mean square error (MSE) satisfies
\begin{equation}
MSE(h) \le C_2^2 h^{2\beta} + \frac {C_1} {Rh}.
\end{equation}
The minimum is obtained for\footnote{Notice that, the minimum $h^{\ast}$ depends on $i$ and $j$ but we chose to omit
the superscripts $i,j$ for the sake of simplicity.}
\begin{equation}
h^* = \left(\frac {C_1} {2\beta C_2^2}\right)^{\frac 1 {2\beta+1}} R^{-\frac 1 {2\beta +1}},
\end{equation}
for which one gets the usual adaptive non-parametric convergence speed
\begin{equation}
\label{MSE}
MSE(h^*) = E({\cal E}_1^2) = O\left( R^{-\frac{2\beta}{2\beta+1}}\right).
\end{equation}
\end{proposition}

\vskip .3cm
Thus, for instance, in the special case where the $\phi^{ij}(t)$'s are H\"older 1 (i.e., $\beta = 1$), using a simple order 0 kernel (such as $K(u) = 1_{[0,1]}(u)$), one gets that if one chooses
$h^*$ of the order of $R^{-1/3}$, we expect the MSE to be of the order of $R^{-2/3}$ where $R$ is the number of iid realizations of the Hawkes process.
Let us point out that, in practice, one often has a single realization.
One can then estimate $g^{ij}+\Lambda^i$ by averaging on the different jumps of this realization.
More precisely, we replace  \eqref{averaged} which used $R$ realizations on the time interval $[0,t_{max}]$
by the following estimation that uses a single realization on the time interval $[0,T]$ (with $T>> t_{max}$) :
\begin{equation}
\label{averaged1}
g_*^{ij}(t) = \frac{1}{Jh} \sum_{n=1}^{J} \int_0^{T} dN^i_u K\left(\frac{u-t^j_n-t}h\right),
\end{equation}
where $J$ is the number of events of the realization.
After a certain time, the realization can be considered as independent of its beginning. One should then get an error which is of the order of
\begin{equation}
\label{e1_}
{\cal E}_1 = O\left( J^{-\frac{\beta}{2\beta+1}}\right),
\end{equation}
so for $\beta = 1$ that gives an error of the order of $J^{-1/3}$.

\begin{figure}[h]
\begin{center}
\includegraphics[height=8cm]{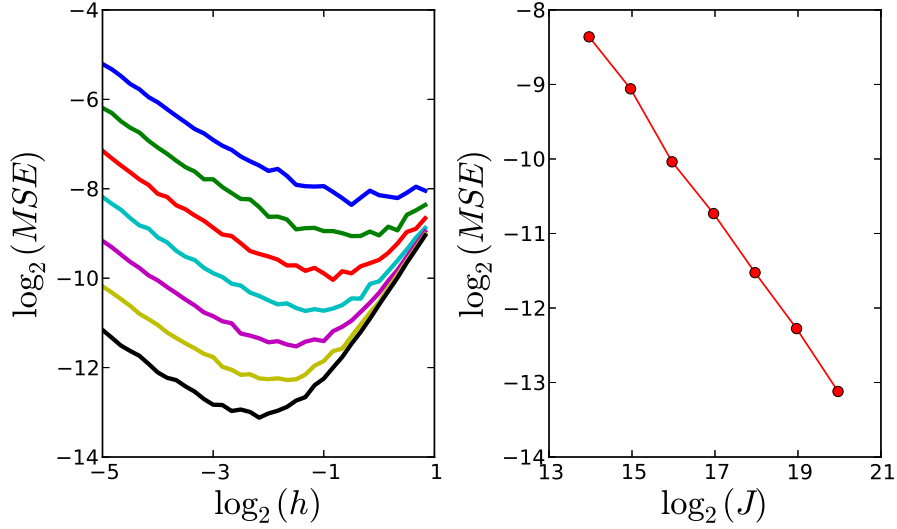}
\end{center}
\caption{Mean integrated square error (MISE) as a function of the kernel width $h$ (left panel)
and the sample size $J$ (right panel) for a 1-dimensional Hawkes model with $\mu = 0.05$ and $\phi(t) = 0.1e^{-0.2 t}$. The errors have been estimated using 500 Monte-Carlo samples of the process. In the left panel are represented the MISE associated
with various sample sizes ($J=8.10^3, 1.6 10^4, 3.2 10^4, \ldots, 1.24. 10^5$). The results are in good agreement with Proposition \ref{propmse}.}
\label{fig_errorvsh}
\end{figure}

These results are illustrated in Fig. \ref{fig_errorvsh} where we have computed the integrated MSE (MISE) on $g(t)$ in a 1-dimensional  Hawkes model with an exponential kernel ($\phi(t) = 0.1 e^{-0.2 t}$, $\mu = 0.05$) for
different realization sizes $J$ increasing from $8. 10^3$
to $10^6$ by a factor 2.
In order to estimate the MISE, for each parameter, we have generated 500 trials of the model and estimated the MISE using the analytical expression of $g(t)$.
One can see in the log-log representation, that for each
sample size ($J$), the error decreases in the small $h$ regime
as expected from Eq. \eqref{mse1} as $h^{-1}$. For large values
of $h$, one clearly observes the bias contribution that is expected to behave like $h^2$. We have checked that the
optimum value ($h^\ast$) behaves for large $J$ as
$h^\ast \simeq C J^{-0.33}$.
In the right panel figure is reported, in log-log scale, the minimum MISE error as a function of the sample size. As predicted by Eq. \eqref{MSE}, one gets a power-law with an
exponent close to $2/3$.

\begin{figure}[h]

\centering
\includegraphics[width=10cm]{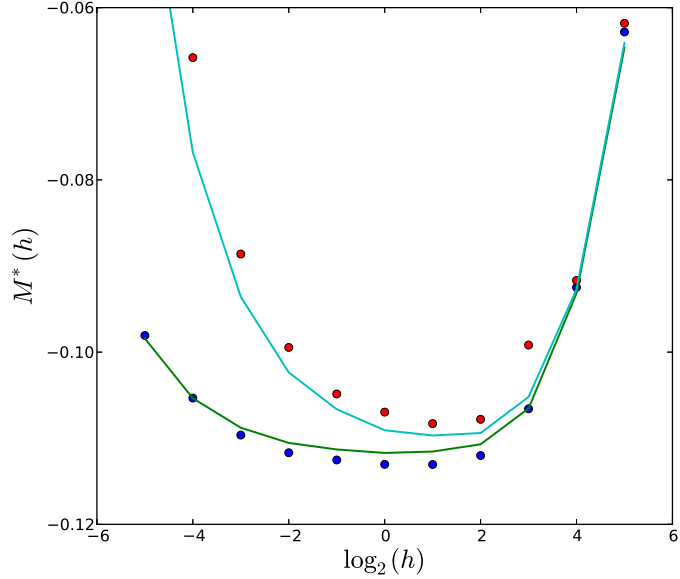}
\caption{Estimation of $M^*(h)$ ($\bullet$) (see \eqref{contrastaverage}) as compared to the theoretical $M(h)$ (solid lines) (see \eqref{MISE}) for a 1-d Hawkes sample. $M(h)$ corresponds to the
$h$ dependent part
of the MISE.
The top lines corresponds to a sample size $J=10^4$ events while
the bottom curves correspond to a sample size $J=5. 10^4$.
 All curves are plotted as a function of $\log_2(h)$.
 In each case the Hawkes model corresponds to the exponential kernel $\phi(t)=0.1 e^{-0.2t}$ with $\mu = 0.05$.
The estimated curves ($\bullet$) are in good agreement with the theoretical curves (solid lines).
The abscissa where the minimum is reached provides an
estimation of optimal bandwidth $h^*$. In both cases we choose $R=10$ in Eq. \eqref{contrastaverage}.
}
\label{fig_contrast}
\end{figure}
\vskip .5cm
\noindent
{\bf Bandwdith selection using cross-validation.}
From a practical point of view, in order to choose the optimal value of the bandwidth $h^{\ast}$ from
a single realization ($R=1$) on the time interval $[0,T]$, one can use a cross-validation method. Indeed, in the MISE computation, the only terms that depend on $h$ read:
\begin{eqnarray}
\label{MISE}
    M(h)   & = & E \left( \int_0^{t_{max}} g_*^{ij}(t)^2 dt - 2 \int_0^{t_{max}} g^{ij}(t) g_*^{ij}(t) dt \right) \\
    \nonumber
       & = & E \left(\int_0^{t_{max}} g_*^{ij}(t)^2 dt  + 2  \Lambda^i \int_0^{t_{max}}  g_*^{ij}(t) dt
       - 2 E\left( \int_0^{t_{max}}  g_*^{ij}(t) dN^i(t) | dN_0^j=1\right)\right)
\end{eqnarray}
where we have considered that
$g^{ij}(t)$ has a support included in $[0,t_{max}]$ and where we
used the definition  $g^{ij}(t) = E(dN_t^i | dN_0^j=1)-\Lambda^i dt$.
Replacing the last term in the expectation of the previous equation by an empirical average,
it results that
the following contrast function provides an (unbiased) estimator of $M(h)$ :
\begin{equation}
\label{contrast2}
 C(h) = \int_0^{t_{max}} [g_*^{ij}(t)^2+2\Lambda^i g_*^{ij}(t) ] dt - \frac{2}{J} \sum_{t_k^j} ~
 \sum_{t_k^j < t_l^i \leq {t_{max}} + t_k^j} g_*^{ij}(t_l^i-t_k^j)
\end{equation}
where $J$ stands for the number of events of $N^{j}$, the first sum is taken on all the jumping times
$t^j_k$ of the component $N^{j}$ and the second sum is taken on all the jumping times
$t^i_l$ of the component $N^{i}$ such that $t_k^j < t_l^i \leq {t_{max}} + t_k^j$.
A cross-validation method can be used to estimate the expectation of $C(h)$. For doing so,
one divides the overall realization time interval $[0,T]$ in $R$ intervals of equal size, i.e.,  $I_r=[(r-1)\frac T {R},r\frac T {R}]$ for $r \in [1,R]$.
Then, for each interval $I_r$, the conditional expectation $g^{ij}$ is estimated following the same ideas as \eqref{averaged1}, averaging only on the jumps of $N^j$ which do not take place in $I_r$. More precisely, we get the estimation $g_{(r)}^{ij}$ defined by
\begin{equation}
\label{averaged2}
g_{(r)}^{ij}(t) = \frac{1}{(J-J_{r})h} \sum_{t_n^j \notin I_r} \int_0^{T} dN^i_u K\left(\frac{u-t^j_n-t}h\right),~~~~\forall t \in [0,t_{max}],
\end{equation}
where $J_r$ corresponds to the number of jumps of $N^j$ that take place in the time-interval $I_r$.
The contrast function \eqref{contrast2} is then estimated on $I_r$, i.e.,
\begin{equation}
\label{contrastn}
 C_{(r)}(h) = \int_{0}^{t_{max}} [g_{(r)}^{ij}(t)^2+2\Lambda^i g_{(r)}^{ij}(t) ] dt
 - \frac{2}{J_r} \sum_{t_k^j \in I_r} ~
 \sum_{t_k^j < t_l^i \leq T + t_k^j} g_{(n)}^{ij}(t_l^i-t_k^j).
\end{equation}
The estimation of $M(h) = E(C(h))$ is thus naturally given by averaging on all the so-obtained quantities for all the possible choices of $n$ :
\begin{equation}
\label{contrastaverage}
 M^{\ast}(h) = \frac{1} R \sum_{r=1}^{R} C_{(r)}(h).
\end{equation}
Examples of $M^{\ast}(h)$ curves are provided in
Fig. \ref{fig_contrast} on a simple 1-d example of Hawkes process. The estimation $M^{\ast}(h)$ is close  the theoretical quantity $M(h)$.
The abscissa corresponding to the minimum of the so-obtained $M^\ast(h)$ curve gives an estimation of the optimal bandwidth $h^*$
$$
h^* = \operatornamewithlimits{\mbox{argmin}}_{h} M^*(h).
$$
\subsection{Overall error - Selection of the number of quadrature points $Q$}
\label{Q}
If one supposes that $g(t)$ is estimated with no error, it remains to study the error due to the quadrature approximation.
The function $K(t,s) = g(t-s)$ is the kernel of the Wiener-Hopf system \eqref{wh}. Even if the $\phi^{ij}(t)$ functions are regular,
this is a singular kernel since it is generally discontinuous on its diagonal  (except if the dimension $D=1$).
There is a huge literature on how to solve numerically Wiener-Hopf systems with a singular kernel on its diagonal (though generally the singularity is much stronger than a "simple" discontinuity). It is out of the scope of this paper to compare all these methods and try to understand which is more appropriate for our estimator.
A pretty popular method is the Nystr\"om method already described above.
It consists in rewriting \eqref{wh} as we did when we rewrote \eqref{system3} into \eqref{system4}, i.e., isolating the singular behavior of the kernel:
\[
g(t) = \Phi(t) + \int \Phi(s) g(t-s) ds = \Phi(t) + \int (\Phi(s) -\Phi(t)) g(t-s) ds + \Phi(t) \int g(t-s)ds .
\]
The last integral term can be estimated "directly", i.e.,  estimating the primitive function of $g(t)$.
The other integral term is approximated
by a quadrature method (we shall use Gaussian quadrature).

Let us suppose again that each element of $\Phi(t)$ is H\"older $\beta$ (we assume here that $\beta \ge 1$) and that they are all bounded by a constant $\phi_\infty$.
Again using \eqref{g} and \eqref{boundpsi},  one can easily show that
each element of $g(t)$ is bounded by :  $g_{\infty} = A_1 \frac{\phi_{\infty}}{1-\rho} (A_2 + \frac{1}{1-\rho})$,
where $A_1$ and $A_2$ are some constants.
Let us also suppose that all the elements of the derivative $\phi'(t)$ are  bounded by a constant  $\phi'_\infty$. Then, the  terms in  $(\Phi(s) -\Phi(t)) g(t-s)$  are of class H\"older $\beta$ everywhere, except on the diagonal $s=t$
for which the derivative might be discontinuous.
If the quadrature method uses $Q$ points and is basically of order greater than $\max(2,\beta)$, the error  for the non diagonal terms is of the order of $\frac{\phi_{\infty}~g_{\infty}}{Q^{\beta}}$ while
the error on the diagonal terms is of the order of
$\frac{{\phi}'_{\infty}~g_{\infty}}{Q^{2}}$.
The total quadrature error is of the order of
\[
\mbox{Quadrature error } \sim  g_{\infty} \left(\frac{\phi_{\infty}}{Q^{2}} + \frac{\phi'_{\infty}}{Q^{\beta}}\right)
 \]
Solving the Wiener-Hopf equation amounts in applying the convolution operator whose Laplace transform is  (see \eqref{lg})
\[
(\Id + \hat g(z))^{-1} =   \Sigma (\Id - \hat \Phi^T(-z))  \Sigma ^{-1} (\Id - \hat \Phi(z)).
\]
This operator has a norm which is bounded by a constant times $\phi_{\infty}$.
Thus the error ${\cal E}_2$ resulting from the Wiener-Hopf inversion is of the order of the quadrature error :
\begin{equation}
\label{E2}
{\cal E}_2  = O\left(Q^{-2} + Q^{-\beta}\right)
\end{equation}

\begin{figure}[h]
\begin{center}
\includegraphics[height=7cm]{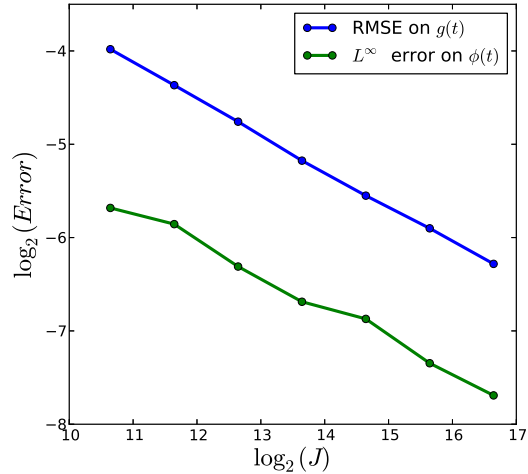}
\end{center}
\caption{Root mean square error on $g(t)$ and $L^\infty$ error on $\phi(t)$ as functions
of the sample size (in log-log scale) for the same Hawkes process as in Fig. \ref{fig_errorvsh}. In both case $h = h^\ast$ and the number of quadradure points $Q$
is kept fixed ($Q = 30$). Both errors behave as $J^{-1/3}$}.
\label{fig_linf}
\end{figure}

Since, as we have just seen, the Wiener-Hopf inversion involves an operator whose norm is bounded by a constant (controlled by $\phi_{\infty}$) the Wiener-Hopf kernel estimation error  ${\cal E}_1$  (related to to error in the estimation of $g(t)$) does not change of magnitude order when inversion is performed.
We deduce that the order of the overall error (when $h = h^\ast$)
should basically be of the order of the sum of the two errors (see \eqref{e1_})
\[
\mbox{overall error  : } {\cal E}_1 + {\cal E}_2 =  O\left( J^{-\frac{\beta}{2\beta+1}}\right) + O\left(Q^{-2} + Q^{-\beta}\right)
\]
Again, as explained in the introduction of this section, we left apart the fact that the inversion of the system involves a random kernel that we need to control in order to control the error.
In that respect, one can always choose $Q$ large enough so that the quadrature error is negligible
as compared to the estimation error (${\cal E}_2 \ll {\cal E}_1$) so that the overall
error will always behave as $O\left( J^{-\frac{\beta}{2\beta+1}}\right)$.

In Fig. \ref{fig_linf}, we have reported the $L^\infty$ error on $\phi(t)$ as a function
of the sample size for the n-dimensional Hawkes process used previously. We set $Q = 30$ and chose, for each $J$, $h = h^\ast$ (i.e. the optimal bandwidth). One can see that the $L^\infty$ error behaves as the the root mean square error (RMSE)  on $g(t)$, i.e.,
as $J^{-1/3}$.
\begin{figure}[h]
\begin{center}
\includegraphics[height=10cm]{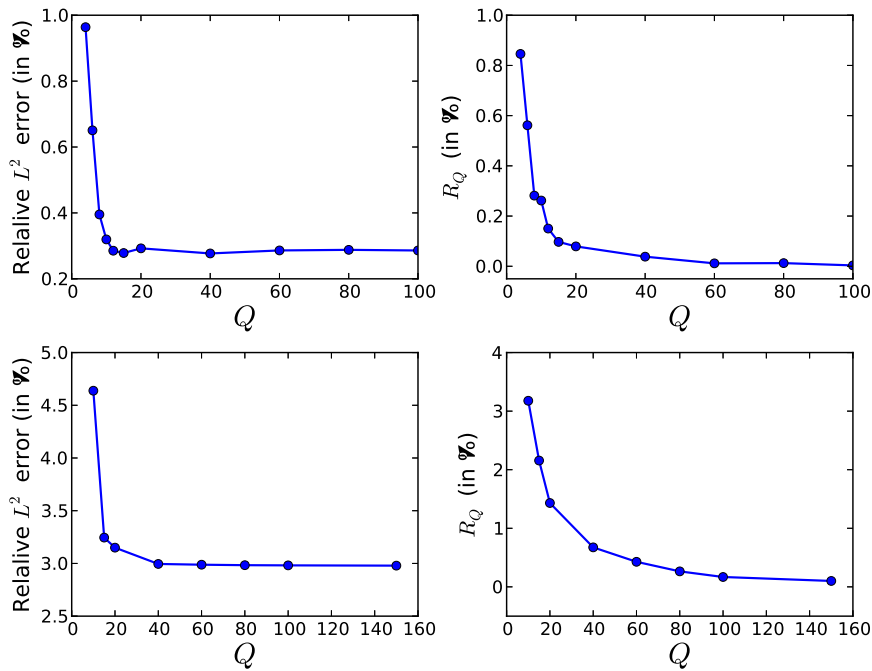}
\end{center}
\caption{Estimation error dependence as a function of the number of quadrature nodes $Q$. Left panels: The relative 
$L^2$ error is displayed as function of $Q$ for a 1D Hawkes process ($J = 10^7$ events) with an exponential 
kernel (top) or power-law kernel (bottom). Right panels: The empirical $L^2$ relative variation $R_Q$ (Eq. \eqref{defRQ}) as a function of $Q$ in the exponential case (top) and power-law case (bottom). One can see that, in both cases, it is sufficient to
choose $Q \in [20,40]$ to be very close to the best estimation for very large $Q$.}
\label{fig_qselection}
\end{figure}

\vskip .5cm
\noindent
{\bf Selection of the number of Quadrature points $Q$.}
From a practical point of view, in order to choose the number $Q$ of quadrature points that are used,
one can start with a relatively small value for $Q$ (e.g. $Q=20$), compute the estimated kernels and compare them to the estimated kernels that are obtained when increasing the value of $Q$. One should stop increasing $Q$ when the two estimations are ``sufficiently close''.
In order to decide whether $Q$ is large enough,
one can e.g. check if the relative $L^2$ variation,
\begin{equation}
\label{defRQ}
R_Q =  \frac {||{\hat \phi}_{Q}-{\hat \phi}_{2Q}||_2}{||{\hat \phi}_{Q}||_2},
\end{equation}
is small enough (for instance $R_Q < 1 \%$).
This is illustrated in Fig. \ref{fig_qselection}, where we have plotted, as a function of $Q$,
both the relative $L^2$ estimation error $||\phi-{\hat \phi}_Q ||_2 / ||\phi||_2$
and $R_Q$ for 2 examples of 1D Hawkes processes involving respectively an exponential (top) 
and power-law kernel (bottom). One can see that in both situations, $Q \in [20,40]$ is large enough.
For all the numerical illustrations and applications presented in this paper, a value in that range has been chosen. Empirically, we observed that, for a wide variety of kernels shapes and for sample sizes between $10^4$ and $10^7$, this range is sufficient to be close to optimal results.




\begin{figure}[h]
\begin{center}
\includegraphics[height=15cm]{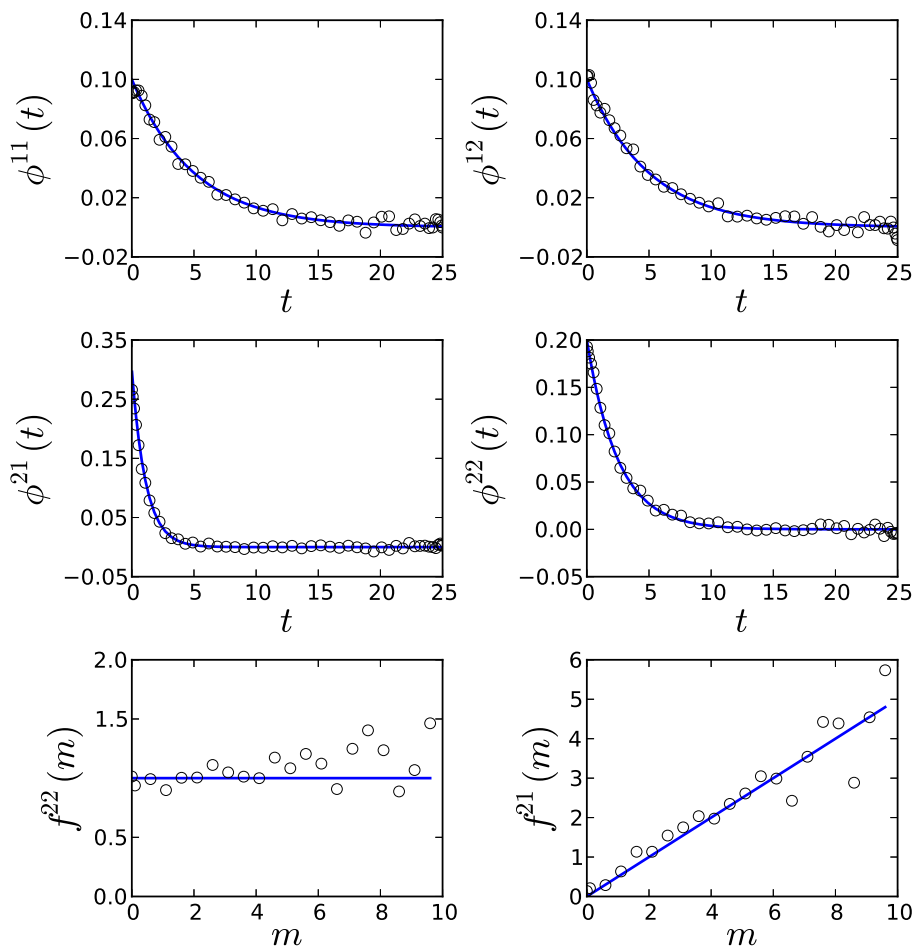}
\end{center}
\caption{Non-parametric estimation of the 2-dimensional marked Hawkes model defined in Section \ref{simu} following estimation procedure of Section \ref{algo}.
Estimation is based on a  realization with  4.5 $10^5$ (resp. 4 $10^5$)
events for component $N^1_t$ (resp. $N^2_t$).
We used $h = 0.5$ and $Q = 50$.
The solid lines represent the analytical curve and the symbols ($\circ$) the empirical estimated values.
The four plots at the top show the kernel matrix $\Phi$ estimation. The two plots at the bottom show the estimation of the mark functions.
See Fig. \ref{fig_qqplot} for goodness of fits.}
\label{fig_estim}
\end{figure}

\begin{figure}[h]
\begin{center}
\includegraphics[height=7cm]{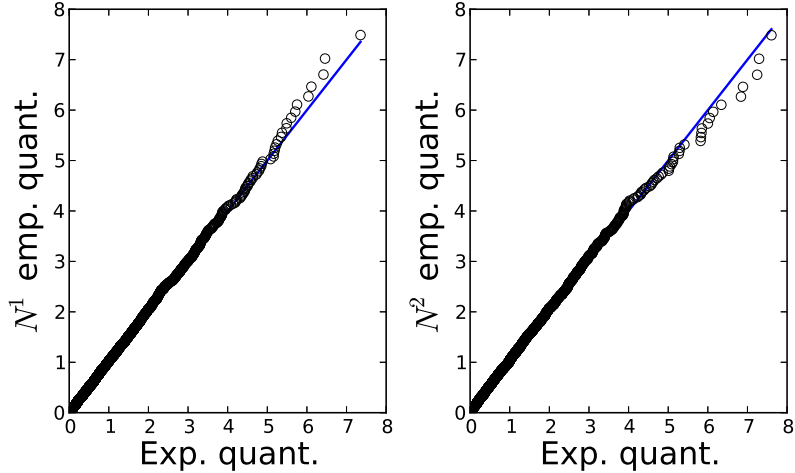}
\end{center}
\caption{Q-Q plots goodness of fits for estimation of the 2-dimensional marked Hawkes model defined in \ref{simu} (see Fig. \ref{fig_estim} for estimation results).
The solid lines represent the expected diagonal and the symbols ($\circ$) the empirical estimated values.}
\label{fig_qqplot}
\end{figure}

\section{Description of the algorithm - Numerical illustrations}
\label{sec:num}
The following section gives a detailed description of the algorithm and the next one gives several numerical illustrations.
\subsection{Description of the algorithm}
\label{algo}
In this Section, we describe precisely each step of our non-parametric estimation procedure within the framework of marked Hawkes processes presented in Section \ref{estimationmethod}.
\begin{itemize}
\item Estimation of the vector $\Lambda$ (simply using as an estimator of $\Lambda^i$ the number of jumps of the realization of $N^i$ divided by the overall time realization)
\item For all $j\in[1,D]$, one must a priori choose $M^j$ and the intervals $\{I^j(l)\}_{1\le l \le M^j}$
\item For all $j \in [1,D]$ and for all $l \in [1,M^j]$, estimation of the probabilities $p^j_l$ as defined by \eqref{plj}. In order to do so, one just needs to count the number of realized marks $\xi_s^j$ that falls within the interval $I^j(l)$.
\item One must estimate $\{G^{ij}_l(nh\}_n$ (on an a priori chosen support $[0,t_{max}]$) using empirical averages as explained in Section \ref{Sec:bandwidth}. For this purpose a value for the bandwidth $h$ must be chosen.
Section \ref{Sec:bandwidth} explains (for the sake of clarity, this section only deals with the simpler case of non-marked Hawkes process, but the generalization is obvious) how this value  can be obtained using cross-validation.
\item Fix the number of quadrature points $Q$ to be used as well as the support for all the kernel functions (we use gaussian quadrature). In general $Q \simeq 30$ is sufficient, however, at the end of Section \ref{Q}, we explain how this value can be chosen adaptively.
Then, for each $i \in [1,D]$, solve the $QM$ linear system obtained by discretizing the convolutions in \eqref{system1} using the quadrature points. This leads to an estimation of all the functions
$\phi^{ij}_l$ on the quadrature points.
\item The estimations of the kernels (on the quadrature points) are simply obtained using the formula
$\phi^{ij}(t) = \sum_{m=1}^{M^j} p_m^j \phi^{ij}_m(t)$. The $L^1$ norm $||\phi^{ij}|| = \int_0^{+\infty} \phi^{ij}(t) dt$ as well as a re-sampling of the kernels on a high resolution grid can be obtained using again quadrature formula.
\item The stability condition of the so-estimated Hawkes process should be checked (i.e., the spectral radius of the $D \times D$ matrix made of the $L^1$ norms $||\phi^{ij}||$ must be strictly smaller than 1).
\item The estimation of the piece-wise values $f^{ij}_l$ of the mark functions $f^{ij}(t)$, can be obtained
via the formula $f^{ij}_l = ||\phi^{ij}_l|| / ||\phi^{ij}||$.
\end{itemize}

\subsection{Numerical illustrations}
\label{simu}
\begin{figure}[h]
\centering
\includegraphics[width=20cm]{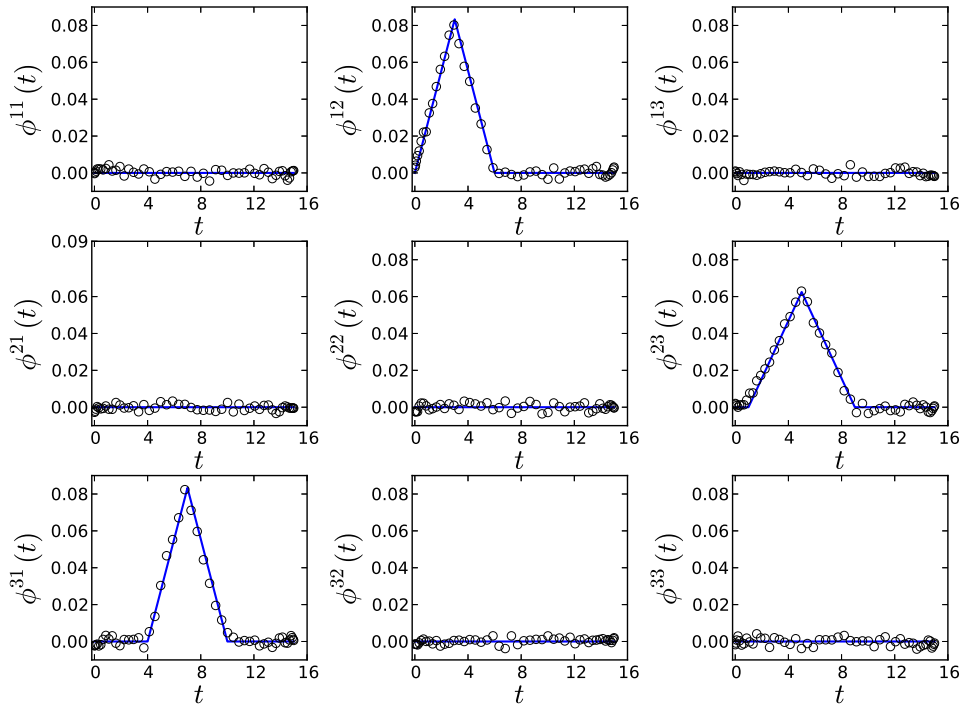}
\caption{Non-parametric estimation of the 3-dimensional Hawkes model with circular interactions.
The estimation is based on a  realization with around $10^5$
events for each component. We used $h = 0.2$ and $Q = 50$.
The solid lines represent the analytical kernels and the symbols ($\circ$) the empirical estimated values.}
\label{fig_estim2}
\end{figure}

\begin{figure}[h]
\centering
\includegraphics[width=12cm]{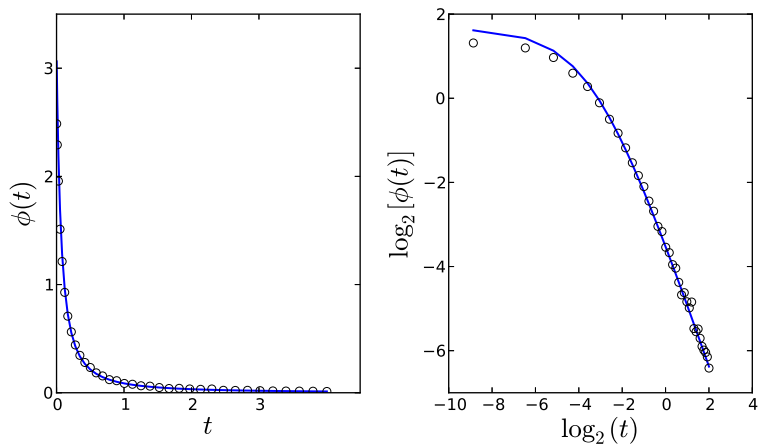}
\caption{Non parametric estimation of the 1-dimensional Hawkes model with a power-law decreasing kernel. The estimation is based on a  realization with around $10^5$
events. We used $h = 0.5$ and $Q = 50$. Left panel: Estimated ($\circ$) and theoretical (solid line) kernels in linear scale. Right panel: Estimated ($\circ$) and theoretical (solid line)
kernels in log-log scale.}
\label{fig_estim3}
\end{figure}

\begin{figure}[h]

\centering
\includegraphics[width=10cm]{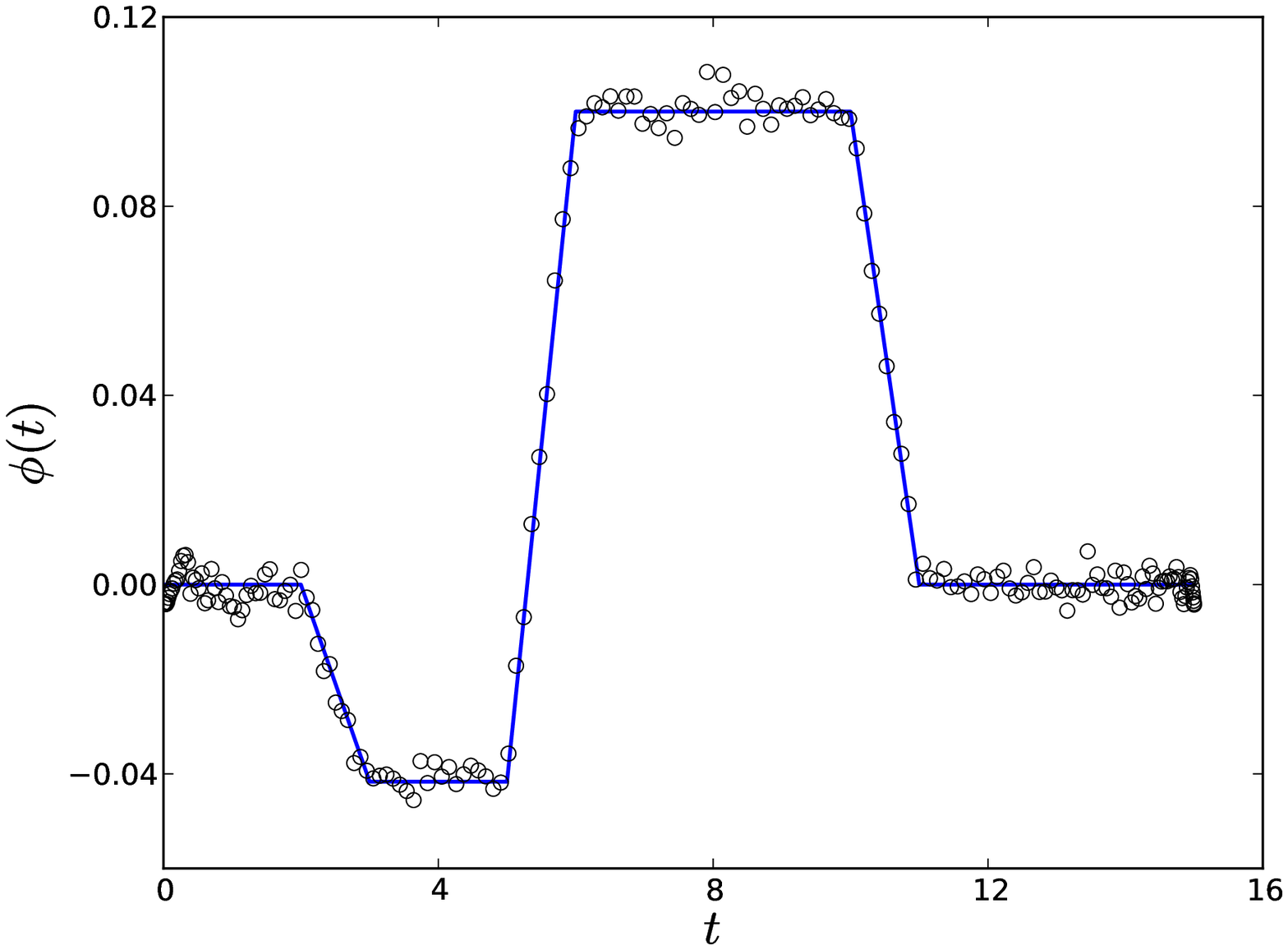}
\caption{Non parametric estimation of a Hawkes model with inhibitory effets. The kernel of the model (solid line) is negative a first interval and positive over another one. If in Eq. \eqref{lambda}, the probability that $\lambda_t <0$ is negligible, the estimation method provides reliable results ($\circ$).}
\label{fig_estim4}
\end{figure}

In this section, we illustrate the performances of the previous estimation method
on various examples accounting for different situations: marked process, non decreasing lagged
kernels, slowly decreasing kernels and kernels that can be negative.

In order to perform numerical simulations of $D$-dimensional Hawkes models, various methods
have been proposed.
We chose to use a thinning algorithm (as proposed, e.g., in  \cite{Ogata1981}). It is an incremental algorithm, the jumping times are generated one after the other.
In its simplest version (in the case of decreasing kernels), at a given time $t$,  it basically consists in
picking up the next potential jumping time $t+\Delta t$ using an exponentially distributed random variable $\Delta t$
with parameter
$\lambda_t = \sum_{k=1}^D \lambda^{k}_t$.
Then a uniform random variable $F$ on $[0,\lambda_t]$ is drawn.
If $F<\lambda_t-\sum_{k=1}^D \lambda^{k}_{t+\Delta t}$,
the jump is  rejected. If not rejected,
 it is assigned
to the $i$th component such that  $i$ is the maximum index $i$
which satisfies $F\ge \lambda_t-\sum_{k=1}^{i-1} \lambda^{k}_{t+\Delta t}$.

\vskip 1cm
\noindent
{\bf A marked 2-dimensional Hawkes process with exponential kernels} \\
The first example we consider is a 2-dimensional marked hawkes process
with exponential kernels. The matrix $\Phi(t)$ of the process reads:
\[
\Phi(t)=
\left( \begin{array}{cc}
0.1 e^{-0.2t} & 0.1 e^{-0.2 t} \\
0.3 e^{-0.9t} & 0.3 e^{-0.4 t}
\end{array} \right)
\]
with $\mu^1 = 0.05$, $\mu^2 = 0.1$. We consider that
$N^2_t$ is marked with a mark $m$ that follows an exponential
distribution of mean value $1$. We choose the mark functions:
$f^{12}(m) = m$ and $f^{22} = 1$, i.e., the mark $m$ only
impacts (linearly) the intensity of the process $N^{1}_t$.

In Fig. \ref{fig_estim} are reported the estimated
kernels and mark functions obtained using a realization
of this 2-dimensional process with  4.5 $10^5$ (resp. 4 $10^5$)
events for component $N^1_t$ (resp. $N^2_t$).
For the mark function estimation, we supposed that $f(m)$ is piecewise constant on intervals $[k/2,(k+1)/2]$ for $k = 0, \ldots, 20$. One can see in the four top figures that each of the exponential kernels
are well estimated (see the next section for a discussion of the error values). In the last two figures, we show that one also recovers the mark functions.

In the estimation procedure, we chose $h = 0.5$ and $Q= 50$ (see Section \ref{error_analysis} for discussion on how to choose these values).

In order to check our estimation and test the Hawkes model on a set of data, one can
perform a goodness-of-fit test by simply noting that
each point process component, $N^{j}(\tilde{t})$, considered as a function of a ``time'' $\tilde{t}  = \int_0^t \lambda^{j}(u) du$, is an homogeneous Poisson process. In that respect the inter events times expressed
in $\tilde{t}$ must be exponentially distributed, i.e., for each $j$, if $t^j_k$ denote de jumping times
of $N^j$, then:
\begin{equation}
\label{tautime}
  \tau^j_k = \int_{t^j_{k_1}}^{t^j_k} \lambda^j(u) du
\end{equation}
must be iid exponential random variables. From the estimated kernels, baseline intensities and mark functions,
one can perform estimations  of all the $\tau^j_k$'s. In Fig. \ref{fig_qqplot} are displayed the Q-Q plots
of the empirical quantiles distribution ($\tau^1_k$ and $\tau^2_k$ were estimated from the last $3.$ $10^3$
intervent times of each components) versus the exponential quantiles. One can check in both case that
both curves are very close to the expected diagonal.

\vskip .7cm
\noindent
{\bf A three dimensional example with circular interactions} \\
In this example, we aim at illustrating two features: first, one can faithfully estimate
Hawkes kernels that are not necessarily decreasing, localized around $t=0$.
One can also, in a multidimensional Hawkes process, disentangle in the complex dynamics of the
events occurrence, the causality (in the sense of Granger causality) relationship between these events.
For that purpose we consider a 3-dimensional Hawkes process with ``circular'' interactions, i.e.,
the process $N^1(t)$ is excited only by the processes $N^2(t)$ itself excited by the process
$N^3(t)$ which events are triggered by the events of $N^1(t)$.
Thus in the Hawkes matrix, only the terms $\phi^{12}(t)$, $\phi^{23}(t)$ and $\phi^{31}(t)$ are
non zero. We choose the shape of these kernels to be triangular functions with different positions.
In Fig. \ref{fig_estim2}, we report the results of the estimation of such process on a sample
where each process has around $10^5$ events.
we can see that one remarkably recovers both the causality relationship between the 3 processes
and the triangular shapes of the interaction kernels.

\vskip .7cm
\noindent
{\bf Kernel with heavy tail} \\
In Fig. \ref{fig_estim3}, is reported the estimation of a 1D Hawkes model with
the power-law decreasing kernel $\phi(t) = \alpha(\nu+t)^{-\beta}$ with $\alpha = \nu = 0.1$
and $\beta = \frac{3}{2}$. Unlike the previous examples, this kernel has an algebraic
decay for large $t$. One can see in both panels of Fig. \ref{fig_estim3} (corresponding
to linear and log-log representation) that, for a sample of $10^5$ events, our estimation is
very close to the expected kernel on a wide range to time scales. Let us notice that in order
to estimate a kernel that is slowly decaying over a support extending over several decades,
the regular sampling schemes of $g(t)$ and $\phi(t)$ is not suited since it would involve
an exponentially large matrix to invert. For that purpose, in ref. \cite{BJM2014}, we propose
a variant of the method of section \ref{estimationmethod} that relies on a different (i.e., logarithmic) time
sampling and that allows one to estimate $g$ when it varies
over a time interval of several orders of magnitude.

\vskip .7cm
\noindent
{\bf Kernel with non-positive values} \\
The last example we provide, concerns a 1D Hawkes process that involves a kernel with negative values. Indeed, according the remark \ref{nlHawkes}, one can consider Hawkes processes with inhibitory
impact of past events (instead of exciting) when the kernel
$\phi(t)$ takes some negative values. In that situation, if, in Eq. \eqref{lambda2}, the events corresponding to $\mu + \int \phi(t-u) dN_u < 0$ occur with a negligible probability, the model remains linear
and one can expect our estimation method to be still reliable. This is illustrated in Fig. \ref{fig_estim4} where
the estimation is performed using a sample (of $10^5$ events)
of a 1D Hawkes model with the kernel represented by the solid line: this kernel is either piecewise constant or piecewise linear and is negative on a whole interval before becoming positive.
According to this model, the occurrence of some event begins by decreasing the probability of occurrence
of further events while, after some time lag, this event has an
opposite impact and increases the process activity.
By comparing estimated and real kernel values, one can see that
even in that case, the method of section \ref{estimationmethod} provides provides
a fairly good estimation of the kernel shape.

\section{Link with other approaches}
\label{otherapproaches}
In the academic literature, there are very few non parametric estimators of the kernel matrix of a Hawkes process.
In the particular case the kernel matrix is known to be symmetric (which is always true if the dimension $D=1$), the method developed  in \cite{BaMuDaEPJB} uses a spectral method
for inverting \eqref{eq:lEdNdNPrime} and deduces an estimation from the second-order statistics. It can be seen as a particularly elegant way of solving the Wiener-Hopf equation when the kernel is symmetric and in that respect,  it is very similar to our approach (though not at all as general of course).

Apart from this method, there are essentially two other approaches for non parametric estimation.

\vskip .2cm
The first one, initiated by \cite{emhawkes}, corresponds to a non parametric EM  estimation algorithm.
It is based on regularization (via $L^2$ penalization) of the method initially introduced by \cite{MaLe10} in the framework of ETAS model for seismology
(see Section \ref{seismic} for ETAS model).
It has been developed for 1-dimensional Hawkes process.
The maximum likelihood estimator is computed using an EM algorithm :
\begin{itemize}
\item The $E$ step basically corresponds to
computing, for all $n$ and $m$,  the probability $p_{nm}$ that the $n$th jump $t_n$ has been "initiated"~\footnote{Here we refer to the "branching"  structure of the Hawkes process.} by the
$m$th jump $t_m$ (where $t_m<t_n$), knowing an approximation of the kernel $\phi(t)$ and of $\mu$.
\item And the $M$ step corresponds to estimating $\phi(t)$ and $\mu$ knowing all the $p_{nm}$.
\end{itemize}
In \cite{emhawkes}, some numerical experiments on particular cases are performed successfully even when the exogenous intensity
$\mu$ depends slowly on time : the whole function $\mu(t)$ is estimated along with the kernel $\phi(t)$ with a very good approximation.
However, as emphasized in \cite{emhawkes},
the convergence speed of the EM algorithm drastically decreases when the decay of the kernel $\phi(t)$ is getting slower (e.g., power-law decaying kernel). Equivalently, it also drastically decreases when one increases
the average number of events that occur on an interval of the same size order as the one
of the support of $\phi(t)$ (keeping constant the overall number of events $J$, i.e., shortening the overall realization time $T$). This can be performed, for instance, by simply increasing the baseline $\mu$ (keeping constant $J$).

\begin{figure}[h]

\centering
\includegraphics[width=10cm]{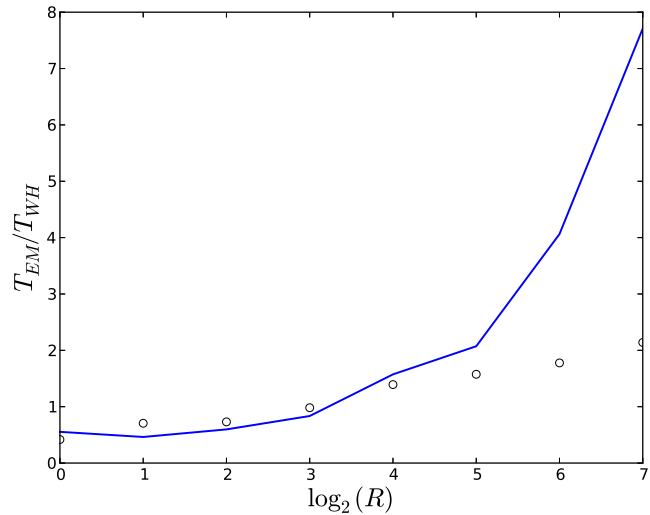}
\caption{ Comparison of Wiener-Hopf and EM complexity: The ratio of the computation time
of EM and Wiener-Hopf methods is plotted as a function of $\log_2(R)$ in the case of a 1d Hawkes model with exponential kernel. For the lower curve (symbol $\circ$) all the parameters are fixed and $R$ represents simply
the ratio of the number of events $J/J_0$ where $J_0 = 10^3$ and $J$ varies from $J_0$ to $2^7 J_0$. On can see that in that case, up to an eventual logarithmic correction, both methods provide a comparable estimation time.
For the upper curve (solid line) the number of events is fixed (to $J = 10^5$ events) while $R$ represents
a normalized average number of events over the support of the kernel $\phi(t)$. More precisely , $R$ is the ratio
$\mu/\mu_0$ where  the baseline rate $\mu$  varies from $\mu = \mu_0$ to $\mu = 2^7 \mu_0$.  One can see that the convergence time of the EM method strongly increases as $\mu$  becomes larger.
}
\label{fig_timeEMvsWH}
\end{figure}

This result is illustrated in Fig. \ref{fig_timeEMvsWH}, where we have compared, for a fixed estimation precision, the computation time $T_{EM}$
of the EM method (without any regularization) and the computation time  $T_{WH}$ of our approach based on solving the Wiener-Hopf equation.
All our tests are using a simple 1d Hawkes process with an exponential kernel.
In a first experiment,
we compared the ratio of the convergence times $T_{EM}/T_{WH}$ as a function of the overall number of events $J$ in the sample.
As illustrated by the curve represented by symbols ($\circ$) in Fig. \ref{fig_timeEMvsWH}, we observed that, up to a logarithmic behavior, the computation times of both methods are comparable. In a second
experiment (solid line curve in Fig. \ref{fig_timeEMvsWH}) we fixed the total number of jumps $J$ and
let the number of events over the support of the kernel $\phi(t)$ vary by simply changing the baseline intensity $\mu$. In that case, one clearly observes that, as $\mu$ increases, the EM approach becomes slower as compared
to our method.

\begin{figure}[h]

\centering
\includegraphics[width=10cm]{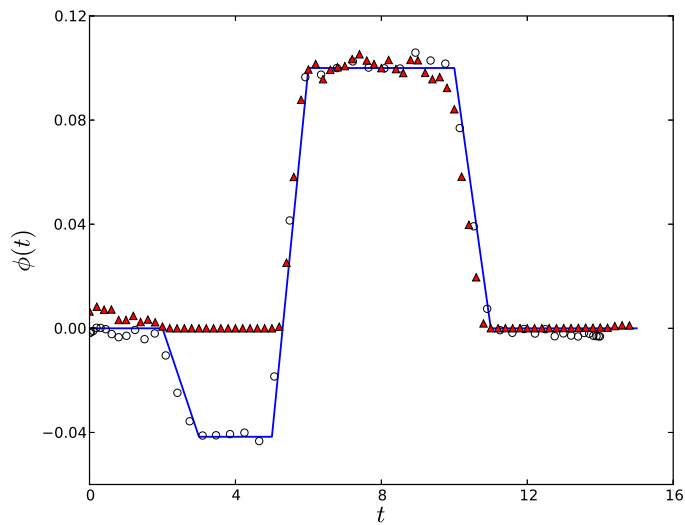}
\caption{Comparison of Wiener-Hopf and EM estimation of the Hawkes model with inhibitory effect of figure \ref{fig_estim4}. Both methods have been tested on the sample sample of size $J \simeq 10^5$ events.
One can see that while our method provides a good estimation of the negative
part of the kernel ($\circ$), the EM based method is only able to estimate its positive part ($\blacktriangle$).}
\label{fig_estimWHEM}
\end{figure}

A second important drawback of the EM approach is illustrated in the figure \ref{fig_estimWHEM}
where we have tested the EM algorithm on the example of Fig. \ref{fig_estim4} where the Hawkes
process involves a kernel with negative values. While our method is able to handle inhibitory
situations (provided the inhibitory effect does not leads to negative intensities), the probabilistic
interpretation of the kernel values involved in the EM method, prevent any such possibility.
One can see in Fig. \ref{fig_estimWHEM} the EM method only allows one to estimate the positive part
of $\phi(t)$.

\vskip .2cm
In a recent series of papers \cite{Rey0,Rey1}, some authors proposed, within a rigourous statistical
framework, a second approach for non-parametric estimation. It  relies on the minimization of the
so-called $L^2$ {\em contrast function}.
Given a realization of a Hawkes process $\tilde N_t$ (associated with the parameters $(\tilde \Phi(t), \tilde \mu)$),
on an interval $[0,T]$, the estimation is based on minimizing the {\em contrast function} $C(\mu,\Phi)$ :
\begin{equation}
\label{contrastmin}
   (\mu^*,\Phi^*) =  \operatornamewithlimits{\mbox{argmin}}_{(\Phi,\mu)} C(\mu,\Phi),
\end{equation}
where
\begin{equation}
\label{contrast}
   C(\mu,\Phi) =   \left( \sum_{i=1}^D \int_0^T \lambda_t^i(\mu,\Phi)^2 dt -  2\int_0^T \lambda_t^i(\mu,\Phi) d\tilde N^i_t  \right)
\end{equation}
and
\begin{equation}
   \lambda_t^i(\mu,\Phi) = \mu^{i} + \sum_{j=1}^D\int_{-\infty}^t \Phi^{ij}(t-s) d\tilde N_s^j \; .
\end{equation}
Let us point out that, minimizing the expectancy of the contrast function is equivalent to minimizing the $L^2$ error on the intensity process. Indeed, if ${\cal F}_t$ is the information available up to time $t^-$, since $E(d\tilde N_t^i | {\cal F}_t) = \tilde \lambda_t^i dt$, one has
\begin{eqnarray*}
\operatornamewithlimits{\mbox{argmin}}_{(\Phi,\mu)}
E(C(\mu,\Phi)) & = & \operatornamewithlimits{\mbox{argmin}}_{(\Phi,\mu)} \sum_{i=1}^D \left( E(\lambda^i(\mu,\Phi)^2)-
E(\lambda^i(\mu,\Phi) \tilde \lambda^i)\right) \\
& = & \operatornamewithlimits{\mbox{argmin}}_{(\Phi,\mu)} \sum_{i=1}^D E\left((\lambda^i(\mu,\Phi)-\tilde {\lambda^i})^2\right).
\end{eqnarray*}
The minimum (zero) is of course uniquely reached for $\Phi = \tilde \Phi$ and $\mu =\tilde \mu$.
Since $\lambda^i_t$ is expressed linearly in terms of  $\mu^i$ and of the $\{\Phi^{ij}\}_j$, minimizing the $L^2$ error is equivalent to solving
a linear equation, which is nothing but the Wiener Hopf equation \eqref{wh}.
Consequently, minimizing the expectancy of the contrast function is equivalent to solving the Wiener-Hopf equation.

In \cite{Rey1}, the authors chose to decompose $\Phi$ on a finite dimensional-space
(in practice, the space of the constant piece-wise functions)
 and to solve directly the minimization problem \eqref{contrastmin} in that space. For that purpose, in order to regularize the solution (they are essentially working with some applications in mind
for which only a
small amount of data is available, and for which the kernels are known to be well localized), they chose to penalize the minimization  with a Lasso term (which is well known to induce sparsity in the kernels), i.e., the $L^1$ norm of the components of
$\Phi$.
Let us point out, that minimizing the contrast function and minimizing the expectancy of the contrast function are two different stories. The contrast function is stochastic, and nothing guarantees that the associated linear equation is not ill-conditioned.
In \cite{Rey1}, the authors prove that, in the case
\begin{itemize}
\item [(i)] the component of $\Phi$ are picked up from an orthogonal family of functions, and
\item[(ii)] any two components of $\Phi$ are always either equal or orthogonal one to each other,
\end{itemize}
 then the linear equation is invertible, i.e., the associated random Gram matrix is almost surely positive definite.
 In applications (they study real signals from neurobiology \cite{Rey2}), they choose the components of $\Phi$ to be piece-wise constant and give a lot of examples of the resulting estimations.

\vskip .2cm
In the present work, our approach is quite different and is motivated for modeling large amount of data (e.g., earthquakes, financial time-series) which are well known to involve very regular non localized kernels.
We directly solve the linear equation, i.e., the Wiener-Hopf system.
In that case, we also introduce a regularization component by the mean of the choice of the quadrature method.
Using a Gaussian quadrature with $Q$ points, for instance, amounts in considering that the kernel term in the Wiener-Hopf equation is polynomial of order $Q$. Let us point out that though we proved that the Wiener-Hopf equation is invertible, we do not have any result on the
invertibility of the stochastic version of the Wiener-Hopf equation in which the conditional expectations are replaced by empirical averages.
As we will see in the next Sections, in all our applications and simulations, it does not seem to be a problem. This is clearly due to the fact that we always consider that a large amount of data is available (e.g., financial high frequency time-series, earthquake time-series).
It seems that if a very small amount of data is available, both approaches described above should be more appropriate than our approach.
For our approach to work, we  would certainly need
to add a regularization term for inverting the discrete linear system.
This will be addressed in a future work.

\section{Examples of application}
\label{sec:applications}
\subsection{Application to financial time-series}
\begin{figure}[h]
\begin{center}
\includegraphics[height=7cm]{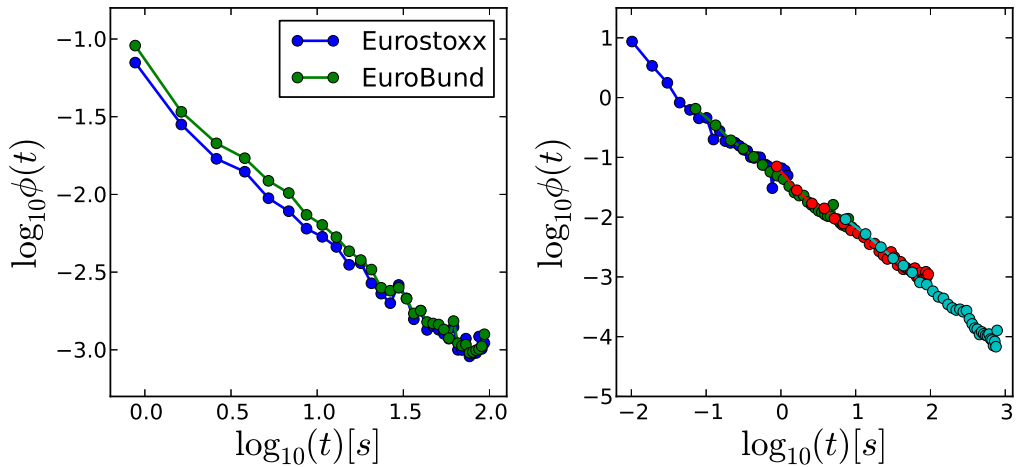}
\end{center}
\caption{Non parametric estimation of a 1-dimensional Hawkes model of market order arrivals
for Eurostoxx (FSXE) and EuroBund (FGBL). (Left) Estimated kernel $\phi(t)$ for FSXE (blue) and FGBL (green) data
in log-log representation. (Right) Estimated kernel $\phi(t)$ for FSXE using over varying time spans
and time resolutions in log-log scale. In both figures the slope is close to $\beta = 1$}
\label{fig_sxegbl}
\end{figure}
Because of their natural ability to account for self and mutual excitation dynamics of specific events, Hawkes processes have sparked an increasing interest in high frequency finance \cite{BH,Hewlett2006Clustering,BDHM2010,Embrechts2011Hawkes,BaMuDaEPJB,BM2013,BouchaudEPJB2013}.
In this section, we consider, as a first application, a 1-dimensional Hawkes model
for market order arrivals. The trading rate in financial markets has non trivial statistical features and is one  the key factors that determines volatility.
We use intraday data of most liquid maturity of EuroStoxx (FSXE) and EuroBund (FGBL) future contracts.
Our data
correspond to all trades at best bid/ask during 1000 trading days covering the period from
may 2009 to september 2013. The number of events (market orders) per day is close to
$5$ $10^4$ for  FSXE and $4.$ $10^4$ for  FGBL.
It is well known that market activity is not stationary and is characterized by a U-like
shape. In order to circumvent this difficulty one can perform an estimation at a fixed
time period within the day. However, as emphasized in  \cite{BM2013}, it appears that the
kernel shape is constant for each period (the non-stationarity of the trade arrival rate
can be associated with a varying baseline intensity $\mu$) and that estimating this kernel on each of these small periods and averaging on all the so-obtained estimations lead
to the same estimation.
This is why we did not consider the problem of intraday non-stationarity and performed
the estimation over a whole trading day.
In left Figure \ref{fig_sxegbl}, we have reported the kernels of trades associated with
FSXE and FGBL as estimated using the method described in section \ref{algo} (without considering any mark).
The curves are displayed in log-log representation since
both kernels are obviously very close to a power-law. They are well fitted by:
\begin{equation}
\label{plkernel}
  \phi(t) = \alpha t^{-\beta}
\end{equation}
with $\beta \simeq 1$ and $\alpha \simeq 0.05$.
In right Figure \ref{fig_sxegbl}, we have reported the same estimation for  FSXE
market orders but with 4 different values of the sampling parameter $h$ (we chose $h = 0.005,0.05,0.5,2$ seconds)
in order to cover a wide range of times $t$. One sees that all
curves consistently fall on the analytical expression \eqref{plkernel}. This figure, where a scaling behavior
of $\phi(t)$ can be observed over a range of 5 decades, is very similar to the estimation performed by Bouchaud {\em et al.} on the S\&P 500 mini \cite{BouchaudEPJB2013}\footnote{It is noteworthy that these authors found values
of $\alpha$ and $\beta$ that are consistent with the previous reported values.}.
Let us notice that origin of the power-law nature of Hawkes kernels
for market data is a challenging question already raised in ref. \cite{BaMuDaEPJB,BM2013}.
It is also remarkable that the shape of the kernel appears to be almost constant for different markets.
Let us finally point out that in order to have $||\phi|| < 1$,
expression \eqref{plkernel} must be truncated at both small
and large time scales. This point is discussed in Refs. \cite{BaMuDaEPJB,BouchaudEPJB2013}. However,
it is important to notice that
the expression \eqref{plkernel} has to be integrated over 9 decades in order to reach $||\phi|| = 1$ (ie., if the
minimum time resolution is $0.001$ s, even after one month, the integral of $\phi$ is still smaller than 1).

\subsection{Application to earthquake time-series}
\label{seismic}
\begin{figure}[h]
\begin{center}
\includegraphics[height=7cm]{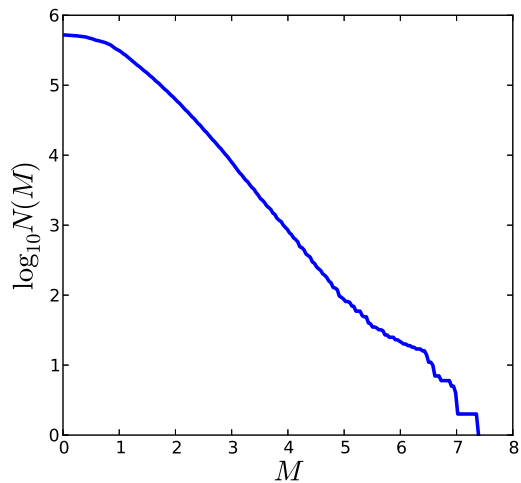}
\end{center}
\caption{
$\log_{10}N(M)$ as a function of $M$ where $N(M)$ is the number of events of magnitude $m\ge M$ as estimated
 from data of the Northern California Earthquake Catalog.
As expected, an exponential shape of the form \eqref{grlaw}, with $M_0 \simeq 2$ and $a \simeq 1$ fits well the empirical
data.}
\label{fig_grlaw}
\end{figure}

Various point processes models have been proposed in order
to reproduce the dynamics of seismic events (earthquakes)
in some given geographic region (see e.g. \cite{ogata99,ppearthquakes} for a review).
Among theses models, the Epidemic Type Aftershock
Sequence (ETAS) proposed by Ogata \cite{Ogata88}
is one of the most popular. This model accounts for the triggering of future
events (aftershocks) by former earthquakes simply by assuming
that the shocks dynamics corresponds to
a Hawkes process marked by the events magnitude.
More precisely, the one dimensional version of this model \footnote{There exists
a space-time version of the ETAS model that accounts for both the time and the location
of earthquakes.} is defined as follows:
$\mu$ is the baseline intensity, the kernel $\phi(t)$ is
\begin{equation}
\label{etasphi}
  \phi(t) = \frac{C}{(1+t/c)^p}
\end{equation}
while the mark function (the ``productivity law'') has an exponential like form:
\begin{equation}
\label{etasf}
  f(m) = A e^{\alpha m } \; .
\end{equation}
Notice that the probability distribution of earthquake magnitudes is
given by the famous Gutenberg-Richer law according to which
the probability that an earthquake magnitude $m$ is greater than $M$ (large enough, i.e., greater than a given threshold $M_0$)
reads
\begin{equation}
\label{grlaw}
  Prob(m>M) = 10^{a-bM} \; ~~\forall \; M \geq M_0
\end{equation}

\begin{figure}[h]
\begin{center}
\includegraphics[height=11cm]{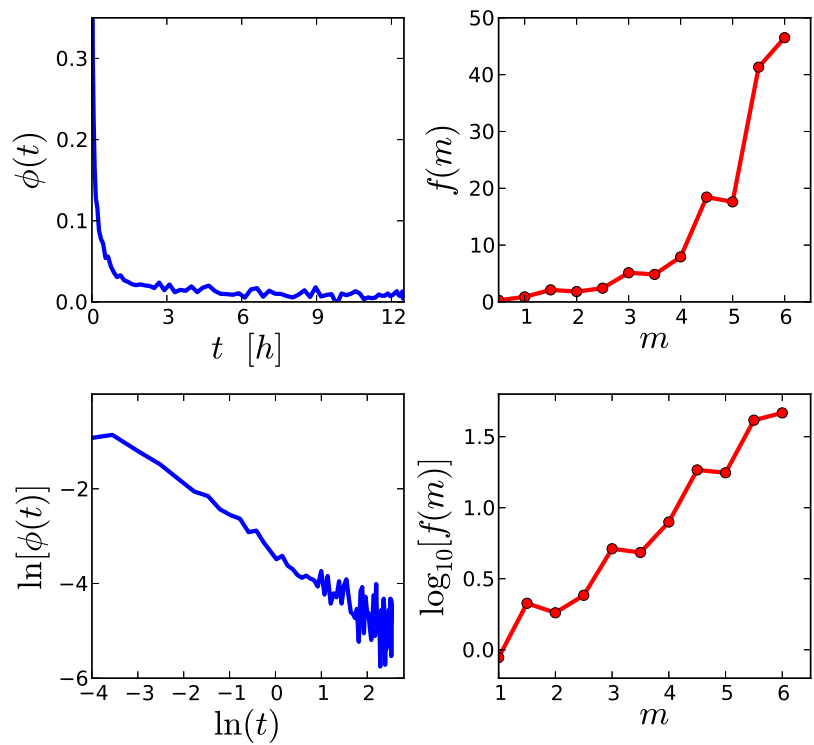}
\end{center}
\caption{Non paramatric estimation of a 1-dimensional marked Hawkes model of earthquake data
from NCEC.}
\label{fig_eq}
\end{figure}

In Fig. \ref{fig_grlaw},  we have displayed $\log_{10}N(M)$ as a function of $M$ where $N(M)$ is the number of
events of magnitude $m\ge M$ as estimated
 from data of the Northern California Earthquake Catalog \cite{ncedc}  (NCEC). This
catalog contains events of magnitude $m \geq 0$ in the region of Northern California
from january 1990 to october 2013. The total number of events is around $5.$ $10^5$.
One can check that the Guttenberg-Richter law \eqref{grlaw} provides a good
fit of the data.
Notice that despite the large number of studies devoted to ETAS model application to
earthquake data (see e.g. \cite{ogata99,ppearthquakes}), to our knowledge, only ref. \cite{MaLe10} performed a non parametric estimate within the general class of linear self-exciting point processes.
As previously mentioned (see section \ref{otherapproaches}) these authors obtained an estimation of
the kernel shape $\phi(t)$ by applying an Expected Maximization iterative method.
A precise comparison of the EM-based method to the one presented in this study
is out of the scope of this paper. Let us just mention that
the authors in \cite{MaLe10} mainly found, using data from southern California Catalog,
a value of the exponent $p$ that weakly depends on the magnitude threshold but stays in the range $p \in [0.6,0.8]$ while the reported value of $\alpha$ in the productivity law \eqref{etasf} is $\alpha \simeq 0.7$.
In Fig. \ref{fig_eq} are reported the results of the non-parametric estimation
of $\phi(t)$ and $f(m)$ from NCEC data. One sees that the obtained estimates for both $f(m)$
and $\phi(t)$ are consistent with ETAS model: in particular one finds
that the kernel $\phi(t)$ is well fitted by the shape \eqref{etasphi}
with $p \simeq 0.7$ and $c \simeq 1$ mn while the law \eqref{etasf}
is well verified over the whole range of magnitudes with $\alpha \simeq 0.75$.
It is noteworthy that the values we obtain are relatively close to the estimates performed
in \cite{MaLe10}.

\section{Conclusion and prospects}
\label{sec:conclusion}
In this paper we have discussed some issues related to Hawkes processes,
an important class of point processes that account for the self and mutual
excitation between classes of events. We have notably
shown that this family of processes are fully characterized by their
conditional density function (or equivalently  by its jumps self-correlation matrix) which is related to the kernel matrix $\Phi$ through a
system of Wiener-Hopf equations. A numerical inversion of this system by
the means of Gaussian quadratures, provides an efficient method for non-parametric
estimation of Hawkes multivariate models. A simple variant of this method allows one
to also recover the mark functions for marked Hawkes processes.
For H\"older $\beta$ kernels,
the estimation error
is shown to converge as $J^{-1/(2\beta+1)}$ (where $J$ is the number of events) provided
one chooses a number of quadrature points large enough. The two examples from finance
and geophysics we considered, illustrate the reliability of the approach and allow
one to confirm the slowly decaying nature of trading activity impact on future activity
and to recover the specifications of the Ogata ETAS model for earthquakes.

As far as the prospects are concerned, we intend to perform a systematic comparison of
our method performances to former methods based on the
minimization of a cost function. It also remains to base our error analysis on more
solid mathematical and statistical foundations. We can also improve the method
by regularity constraints in the case a very small amount of data is available.
The estimation of the mark functions
also raises interesting problems: for example one could test, within
the framework of this paper, whether the kernels are separable, i.e.,
$\phi(t,m) = \phi(t) f(m)$ or not.

As we mentioned in the introduction, since they naturally and simply capture a causal structure
of event dynamics associated with contagion, cross and self activation phenomena,
Hawkes processes are promised for many applications.
In that respect, we hope that the method proposed in this paper will enter in the toolbox
of standard techniques used for empirical applications of point processes.
Besides an improvement of the applications to finance (e.g. by accounting for order book events, price events or exogenous events) and geophysics (e.g. by accounting for the space dependence of earthquake mutual excitations), we plan to consider other fields where Hawkes processes are pertinent like social networks information diffusion,
internet activity or neural networks

\section*{Acknowledgments}
The authors thank  Fran\c{c}ois Alouges, St\'ephane Gaiffas, Marc Hoffmann, Patricia Reynaud-Bouret, Mathieu Rosenbaum and Vincent Rivoirard for useful discussions.
We gratefully acknowledge financial
support of the chair {\it Financial Risks}
of the {\it Risk Foundation},
of the chair
{\it Mutating Markets} of the French Federation of Banks and
of the chair
QuantValley/Risk Foundation: Quantitative Management Initiative.

The financial data used in this paper  have been  provided by the company {\em QuantHouse
EUROPE/ASIA}, http://www.quanthouse.com.

We also aknowledge the Northern California Earthquake Data Center (NCEDC),
Northern California Seismic Network, U.S. Geological Survey, Menlo Park
Berkeley Seismological Laboratory, University of California, Berkeley.

\appendices
\section{Proof of the unicity of the solution of the system \eqref{wh} in $\Phi(t)$}
\label{sec:unicity}
Let us show that we can use a standard Wiener-Hopf factorization argument to prove that equation \eqref{wh} admits a unique solution matrix $\Phi(t)$ which components are $L^1$ causal functions.
For that purpose let us suppose that $\tilde \Phi(t)$ is another causal solution and let us consider the matrix:
\[
  \Delta(t) = \Phi(t) - \tilde \Phi(t) \; .
\]
We want to prove that $ \Delta(t)$ is equal to 0 for all $t$.

\noindent
Let us set
\begin{equation}
\label{whtrick}
 B(t) =   \Delta(t) + \Delta \star g(t) \; .
\end{equation}
Since both $\Phi(t)$ and $\tilde \Phi(t)$ satisfy \eqref{wh},
one has:
\[
  B(t) = 0 \; \; \forall t > 0,
\]
consequently $B(t)$ is  in $L^1$ and anti-causal.
In the Laplace domain, \eqref{whtrick} writes
\[
  \hat B(z) = \hat \Delta(z) \left(\Id + \hat g(z) \right)
\]
Thanks to  \eqref{lg} and \eqref{lpsi}, one  has:
\begin{eqnarray*}
   \hat B(z) & = & \hat \Delta(z) \left(\Id + \hat \Psi(z) \right) \Sigma \left(\Id+\hat \Psi^T(-z) \right) \Sigma^{-1} \\
   & =& \hat \Delta(z) \left(\Id + \hat \Psi(z) \right) \Sigma \left(\Id-\hat \Phi^T(-z) \right)^{-1} \Sigma^{-1}
\end{eqnarray*}
and consequently
\begin{equation}
\label{analytic}
\hat B(z) \Sigma \left(\Id-\hat \Phi^T(-z) \right) =  \hat \Delta(z) \left(\Id + \hat \Psi(z) \right) \Sigma
\end{equation}
On the one hand, since the matrices
$\Psi(t)$ and $\Delta(t)$ are causal and in $L^1$,  the left hand side of this last equation is analytic
in the left half-plane $\{z, \Re(z) \leq 0 \}$.
On the other hand, since the  the matrices
$B(t)$ and $\Phi(-t)$ (the Laplace transform of $\Phi(-t)$ is $\hat \Phi(-z)$) are in $L^1$ and anti-causal, the right hand side
of this last equation is analytic in the right half-plane
$\{z, \Re(z) \geq 0 \}$. Since both sides are equal, they actually are analytic on the whole complex plane (i.e., their elements are entire functions).

\noindent
Let us fix $0< \beta < 1$ and $i,j \in [1,D]$.
We choose $z\in \mathbb{C}$ such that $\Re(z) = -r \in \mathbb{R}^{-*}$.
Then for any causal function $f(t) \in L^1$, one has
\begin{eqnarray*}
\hat f(z) \leq  \int_0^\infty |f(t)| e^{-tr} dt & = & \int_0^{r^{-\beta}} |f(t)| e^{-tr} dt + \int_{r^{-\beta}}^{\infty} |f(t)| e^{-tr} dt
 \leq  \int_0^{r^{-\beta}} |f(t)| dt + e^{-r^{1-\beta}}||f||_1
  \operatornamewithlimits{\longrightarrow}_{r\rightarrow +\infty} 0
\end{eqnarray*}
Since both $\Delta(t)$ and $\Psi(t)$ are causal and in $L^1$, the analytic function of \eqref{analytic} goes to 0 when $\Re(z) \rightarrow -\infty$.
By Liouville theorem, we conclude that it is zero everywhere and consequently $\forall t \ \mathbb{R},~\Delta(t) = 0$.
This ends the proof of the uniqueness of the solution of \eqref{wh} :

\section{Proof of proposition \ref{prop:inteqmarked}}
\label{app:proofinteqmarked}
For a fix $t > 0$, one has, from definition \eqref{defGij}:
\begin{equation}
G^{ij}(t,x)dt dx= E(\lambda^i_t dt~ |~ dN^j_0=1,~\xi^j_0 \in [x,x+dx])  -\Lambda^i dt dx.
\end{equation}
Using \eqref{mlambdai}, we get
\begin{equation}
G^{ij}(t,x)dx = (\mu^i -\Lambda^i)dx + \sum_{k=1}^D \int \phi^{ik}(t-s)E(f^{ik}(\xi^k_s)dN^k_s~ |~ dN^j_0=1,~\xi^j_0 \in [x,x+dx]).
\end{equation}
Then using \eqref{Lambda}, one gets $(\mu^i -\Lambda^i) = - \Lambda^k \sum_{k=1}^D \int ds \phi^{ik} (t-s)$ thus
\begin{equation}
G^{ij}(t,x)dx = \sum_{k=1}^D \int \phi^{ik}(t-s)\left(E(f^{ik}(\xi^k_s)dN^k_s~ |~ dN^j_0=1,~\xi^j_0 \in [x,x+dx])-\Lambda^kdsdx\right).
\end{equation}
Splitting the integral in 3 parts : $s=0$, $s>0$ and $s<0$, we get
\begin{equation}
G^{ij}(t,x)  =
 \phi^{ij}(t)f^{ij}(x) + A^+(t,x) + A^-(t)
\end{equation}
where
\begin{equation}
A^+(t,x)dx = \sum_{k=1}^D \int_{s>0} \phi^{ik}(t-s)\left(E(dN^k_s~ |~ dN^j_0=1,~\xi^j_0 \in [x,x+dx])-\Lambda^k dsdx\right) =
\sum_{k=1}^D \int_{s>0} \phi^{ik}(t-s) G^{kj}(s,x)dx
\end{equation}
and
\begin{equation}
A^-(t) =  \sum_{k=1}^D \int_{s<0} \phi^{ik}(t-s)\left(E(f^{ik}(\xi^k_s)dN^k_s~ |~ dN^j_0=1)-\Lambda^k ds\right).
\end{equation}
Since, for $s<0$
\begin{eqnarray*}
E(f^{ik}(\xi^k_s)dN^k_s~ |~ dN^j_0=1)
& = &
\int dz f^{ik}(z) \frac {Prob\{ \xi^k_s = z, dN^k_s = 1, dN^j_0 = 1\}}
{Prob\{dN^j_0 = 1\}} \\
& = &
\int dz  f^{ik}(z) \frac {Prob\{ \xi^k_s = z, dN^k_s = 1, dN^j_0 = 1\}}
{\Lambda^j} \\
& = &
\frac{\Lambda^k }{\Lambda^j} \int dz  f^{ik}(z)
p^k(z)
E(dN^j_0 ~|~\xi^k_s=z,dN^k_s =1) \\
& = &
\frac{\Lambda^k }{\Lambda^j} \int dz  f^{ik}(z)
p^k(z) \left( G^{jk}(-s,z) + \Lambda^j\right).
\end{eqnarray*}
Consequently
\begin{equation}
A^-(t) =  \sum_{k=1}^D \frac{\Lambda^k }{\Lambda^j}  \int_{s<0} \phi^{ik}(t-s)\
\int dz  f^{ik}(z)
p^k(z) G^{jk}(-s,z)
\end{equation}
We finally get
\begin{equation}
G^{ij}(t,x)  =
 \phi^{ij}(t)f^{ij}(x) + \sum_{k=1}^D \int_{s>0} \phi^{ik}(t-s) G^{kj}(s,x) +
 \sum_{k=1}^D \frac{\Lambda^k }{\Lambda^j}  \int_{s<0} \phi^{ik}(t-s)\
\int dz  f^{ik}(z)
p^k(z) G^{jk}(-s,z)
\end{equation}
This proves proposition \ref{prop:inteqmarked}.

\section{Proof of proposition \ref{propmse}}
\label{app:esterror}
The proof of proposition \ref{propmse} follows a path that is very
classical in the study of density kernel estimations or regressions.
First, the variance of $g_*(t)$ can be bounded
\begin{eqnarray}
Var(g_*(t)) & \le & \frac{1}{R^2h^2} E\left(\left(\sum_{k=1}^{R} \int_0^{t_{max}} dN^i_u(k)  K\left(\frac{u-t^j_1(k)-t}h\right)\right)^2\right) \\
& \le & \frac{1}{Rh^2} E\left(\left( \int_0^{t_{max}} dN^i_u K\left(\frac{u-t^j_1-t}h\right)\right)^2\right).
\end{eqnarray}
Using the stationarity of the increments of $N_t$
\begin{eqnarray}
Var(g_*(t)) & \le &  \frac{1}{Rh^2} E\left(\left( \int_0^{t_{max}} dN^i_u K\left(\frac{u-t}h\right)\right)^2 | ~ dN_0^j = 1 \right) \\
\label{dom}
& \le  &  \frac{1}{Rh^2} \int_{u \in ]0,{t_{max}}]} \int_{v \in ]0,{t_{max}}]} E(dN^i_u dN^i_v | dN_0^j = 1) K\left(\frac{u-t}h\right)K\left(\frac{v-t}h \right).
\end{eqnarray}
Straightforward computations (see \cite{BaMuDaEPJB} for examples) show that,
$E(dN^i_u dN^i_v | dN_0^j = 1)$ can be decomposed in the sum $\psi^{ij}(u) \delta(u-v) + b(u,v)dudv$, where $\delta(.)$ is the Dirac distribution and $b(u,v)$ is a  polynomial in $\psi^{ij}$ taken at different points.

Let us consider that all the $\phi^{ij}(t)$'s are bounded by a constant $\phi_{\infty}$.
Then, if $F(t)$ is a matrix function with positive elements, one can easily check that each element of the matrix $\Phi \star F(t)$ are
bounded by $\phi_{\infty} {\bf 1} ||F||$ where ${\bf 1}$ is the matrix whose elements are all equal to 1 and $||F||$ the matrix made of the
$L^1$ norm of the elements of $F(t)$. Then applying this result to $F(t) = \sum_{k=0}^{+\infty} \Phi^{(*k)}(t)$, and $\Psi(t) = \Phi \star F(t)$,
shows that
\begin{equation}
\label{boundpsi}
\mbox{each element of $\Psi(t)$ is bounded by : } \psi_{\infty} = \phi_{\infty} {\bf 1} (\Id-||\Phi||)^{-1} < C  \frac{\phi_{\infty}}{1-\rho}
\end{equation}
where $C$ is a constant and $\rho$ the spectral radius of $\Id-||\Phi||$.
Consequently, since (H) is supposed to hold (i.e., $\rho<1$), when $h$ is small, the dominant term
in \eqref{dom}
is of the form
$$
\frac{1}{Rh^2} \int_{u \in ]0,{t_{max}}]} \int_{v \in ]0,{t_{max}}]}
\psi^{ij}(u)\delta(u-v)
 K\left(\frac{u-t}h\right)K\left(\frac{v-t}h \right) \le
\frac{\psi_{\infty}}{Rh^2} \int_{u \in ]0,{t_{max}}]}
 K\left(\frac{u-t}h\right)^2,
$$
which is of order $1/h$.
Thus, for $h$ small enough, there exists a constant $C_1$ such that
\begin{equation}
\label{mse1}
Var(g_*(t)) \le  \frac{C_1}{Rh}
\end{equation}

For the bias, the computation is also standard :
\begin{eqnarray}
b(t) = E(g_*(t))-g^{ij}(t)-\Lambda^i & = &  \frac{1}{h}E\left(\int_0^{t_{max}} dN^i_u(k) K\left(\frac{u-t^j_1(k)-t}h\right)\right) - g^{ij}(t)-\Lambda^i \\
& = &  \frac{1}{h}\int_0^{t_{max}} E(dN^i_u | dN^j_0 = 0) K\left(\frac{u-t}h\right) - g^{ij}(t)-\Lambda^i \\
& = &  \frac{1}{h}\int_0^{t_{max}} E(dN^i_u | dN^j_0 = 0) K \left(\frac{u-t}h\right) - g^{ij}(t)-\Lambda^i \\
& = &   \int_0^{t_{max}} E(dN^i_{t+uh} | dN^j_0 = 0) K (u) - g^{ij}(t)-\Lambda^i \\
& = &  \int_0^{t_{max}} (E(dN^i_{t+uh} | dN^j_0 = 0) - E(dN^i_{t} | dN^j_0 = 0) du) K (u) \\
& = &  \int_0^{t_{max}} (g^{ij}(t+uh)-g^{ij}(t)) K (u) du
\end{eqnarray}

If follows that if the density of $E(dN^i_{t} | dN^j_0 = 0))$ is H\"older $\beta$ and if the kernel $K$ is of order $l = \lfloor \beta -\epsilon\rfloor$, for all small $\epsilon > 0$~\footnote{The use of $\epsilon$ here is just to take car of the case the $\phi^{ij}$'s are H\"older $n$ with $n$ being an integer}
(i.e., $l$
is the largest integer smaller than $\beta-\epsilon$)
then, there exists a constant $C_2$ such that
\begin{equation}
\label{b1}
|b(t)| \le C_2 h^{\beta}
\end{equation}

The proposition directly results from Eqs \eqref{mse1} and \eqref{b1}.

\bibliography{hestim2}

\end{document}